\documentclass[a4paper]{article}
\usepackage{RR}
\usepackage{hyperref}
\usepackage{mytex}

\newcommand{\jac}{{\cal J }}

\newcommand{\intint}{\int\hspace{-2pt}\int}
\newcommand{\spfds }[1]{<\hspace{-3pt}<\hspace{-2pt} #1 \hspace{-2pt} >\hspace{-3pt}>}
\newcommand{\spm}[1]{<\hspace{-2pt} #1 \hspace{-2pt} >}
\newcommand{\beqnn}{$$}
\newcommand{\eeqnn}{$$}
\RRdate{Novembre 2008}

\RRauthor{
Eugene Kazantsev\thanks{INRIA, 
 Laboratoire Jean Kuntzmann,
BP 53
38041 Grenoble Cedex 9 
France (Eugene.Kazantsev@imag.fr)}%
}
\authorhead{E. Kazantsev}
\RRtitle{Sensibilit\'e d'un mod\`ele barotrope de l'oc\'ean aux perturbations de topographie. }

\RRetitle{Sensitivity of a Barotropic Ocean Model to Perturbations of the Bottom Topography.}

\titlehead{Sensitivity of a Barotropic  Model to  the Bottom Topography.}
\RRresume{  Dans ce papier nous nous int\'eressons \`a l'op\'erateur qui d\'efinit le lien entre les petites erreurs dans la repr\'esentation de la topographie du fond d'un mod\`ele barotrope et sa solution. L'\'etude montre que la solution est tr\`es sensible aux perturbations de la topographie dans les r\'egions o\`u le flux est turbulent. D'un autre cot\'e, le flux est peu sensible dans les r\'egions o\`u l'\'ecoulement est laminaire. La mesure quantitative de la sensibilit\'e est surtout influenc\'ee par le temps de croissance de l'erreur. Sur de courtes \'echelles de temps, la d\'ependance de la sensibilit\'e au temps de croissance de l'erreur est lin\'eaire, mais sur les grandes \'echelles cette d\'ependance devient exponentielle.   
}
\RRabstract{In this paper, we  look for an operator that describes the relationship between  small errors in  representation of the bottom topography in a barotropic  ocean model and the model's solution.  The study shows that the model's solution is very sensitive to topography perturbations in  regions where the flow is turbulent. On the other hand, the flow exhibits  low sensitivity in laminar regions.  The quantitative measure of sensitivity is  influenced  essentially by the error growing time. At short time scales, the sensitivity exhibits the polynomial dependence on the error growing time. And in the long time limit, the dependence becomes exponential. }
\RRmotcle{Sensibilit\'e, Mod\`ele barotrope, Topographie de fond.}
\RRkeyword{Sensitivity, Bottom topography, Barotropic model}
\RRprojets{MOISE}
\RRtheme{ \THNum } 
 \URRhoneAlpes 

\begin{document}
\RRNo{6717}
\makeRR 

\bibliographystyle{plain}

\section{Introduction.}

It is well known the solution of a numerical model depends on a number of parameters and parametrisations. One of them,  bottom topography, attracts much attention because  it plays an important role in the determining the flow field in the ocean (see for example \cite{Holland73},\cite{Adcroft97}). However,  even if the topography  is well described, it is not evident how to represent it in the model on the model's grid because of the limited resolution. It is known for 30 years, that requiring the large scale ocean flow to be well represented, one have to smooth the topography to get only corresponding large-scale components of relief  \cite{Ilin}. In this case, the influence of subgrid-scales has to be parameterized. But it is not evident how to do that, and how to apply the parametrisation for a given model with a given resolution. 
It is shown in  \cite{Penduff02}, that different smoothing of the topography pattern may significantly  change the model's properties. 

One of possible ways to adapt  real bathymetry to a particular model is to perform a data assimilation procedure with the topography as the control parameter. The control parameter in this procedure is  supposed to be modified  to bring the model within an estimated error of the observations. This idea has already been applied in \cite{LoschWunsch} for a linear shallow-water model in a zonal channel. 

In order to proceed to  data assimilation with non-stationary solution of a  nonlinear   model, we evaluate first the sensitivity of  model to the topography variations. In this paper we address especially the most sensitive and the most insensitive modes of the solution with respect to little variations of the topography.  In particular, we look for  modes to which the solution is not sensitive at all. The presence of such  modes indicates the impossibility to  reconstruct the bottom topography from observations in unique way. If the solution exhibits no sensitivity to some mode, then  the topography may be perturbed by adding this mode with no flow change. Mathematically speaking, this mode belongs to the null space of the sensitivity operator. The dimension of the null space determines the  number of independent topography variations resulting in the same observable flow. The existence of the null space has recently been pointed out by \cite{LoschWunsch} for a shallow water-model on the C grid.   In this paper we also discuss the null space of the operator that describes the model's sensitivity.  

 On the other hand, the most sensitive modes will form the sensitive  space. Any small perturbation of the topography by a function from this space will result in a drastic change of the flow. Concerning the data assimilation procedure, it is this space that has to be assimilated in the best and in the fastest way. The dimension of this space shows the least number of functions participating in the cost functional.  

The model used in this paper is a simple barotropic vorticity equation over topography. This model is used in two configurations: a square with  flat bottom and the North Atlantic region with realistic topography. We discuss the steady state solution as well as a non-stationary flow. The sensitivity of an ocean general circulation model to bottom topography has already been studied by adjoint method in \cite{LoschHeimbach}, but  using  a simple well known model allows us to perform a complete study in the whole phase space. In particular, a simplest model makes it possible  to compare  the sensitivity to the topography perturbations with the sensitivity to perturbations of other model's parameters, like it's initial conditions.  

 The paper is organized as follows. The second section describes the model and the sensitivity estimates. In the third section we present the numerical discretisation of the problem.  The fourth and the fifth sections are devoted to experiments in the square box and in the North Atlantic respectively.

\section{Sensitivity estimates.}

We consider  shallow-water model with the rigid lid assumption
\beqr
\der{u}{t} - fv + u\der{u}{x}+v\der{u}{y}  +\fr{1}{\rho_0} \der{p}{x} &=& \frac{\tau^{(x)}}{\rho_0 H_0} -\sigma u + \nu \Delta u\nonumber\\
\der{v}{t} + fu +  u\der{v}{x}+v\der{v}{y} +\fr{1}{\rho_0}\der{p}{y} &=& \frac{\tau^{(y)}}{\rho_0 H_0} -\sigma v + \nu \Delta v \label{sw}\\
\der{(H u)}{x} + \der{(H v)}{y} &=& 0 \nonumber
\eeqr
where $\rho_0$ is the mean density of water and $H_0$ is the characteristic depth of the basin. The Coriolis parameter $f$ is supposed to be linear in $y$ coordinate: $f=f_0+\beta y$.

The third equation allows us to introduce the streamfunction $\psi$, such as 
\beq
 Hu=-\der{\psi}{y},\hspace{5mm} Hv=\der{\psi}{x} 
\eeq
Denoting the vorticity by $\omega=\der{v}{x}-\der{u}{y}$ we get 
\beq
\omega =  \der{}{x} \fr{1}{H} \der{\psi}{x} + \der{}{y} \fr{1}{H}\der{\psi}{y} \label{stat}
\eeq 
Using this notation we calculate the curl of the first two equations of the system~\rf{sw}. We get 
\beq
\der{\omega}{t} +(\omega+f)\; div\; \vec{u} +u\der{(\omega+f)}{x}+v\der{(\omega+f)}{y} = \nu\Delta\omega -\sigma\omega +  \frac{{\cal F}(x,y) }{\rho_0 H_0} \nonumber 
\eeq
or 
\beq
\der{\omega}{t} +\jac(\psi, \fr{\omega+f}{H} )= \nu\Delta\omega -\sigma\omega + \frac{{\cal F}(x,y) }{\rho_0 H_0} \label{btp} 
\eeq
where $\jac(\psi,\omega)=\der{\psi}{x}\der{\omega}{y}-\der{\psi}{y}\der{\omega}{x}$ is the Jacobian operator and ${\cal F}(x,y)=-\fr{\partial \tau_x}{\partial y} +
 \fr{\partial \tau_y}{\partial x}$.  

The system \rf{btp} is considered in the bounded domain $\Omega$ and is subjected to the impermeability and slip boundary conditions:
\beq 
\psi\mid_{\partial\Omega} =0,\;\omega\mid_{\partial\Omega} =0 \label{bcref}
\eeq 

Let us suppose the couple $\psi(x,y,t),\omega(x,y,t)$ is a solution of the system \rf{stat}, \rf{btp} with a given topography $H= H(x,y)$. If we perturb the topography by some small $\delta H$, we get another solution of the system  $\{\psi+\delta\psi, \omega+\delta\omega\}$. 

Our purpose is to define the relationship between $\delta H$ and $\delta\omega$ supposing both of them to be sufficiently small: 
$$\norme{\delta H} \ll \norme{H} \mbox{ and } \norme{\delta \omega} \ll \norme{\omega}$$ 

We start from the stationary equation \rf{stat}. So far, the perturbed couple $\{\psi+\delta\psi, \omega+\delta\omega\}$ is  the solution of the system with  perturbed topography, it must also satisfy the equation \rf{stat}  
\beq
\omega+\delta\omega =  \der{}{x} \fr{1}{H+\delta H} \der{\psi+\delta\psi}{x} + \der{}{y} \fr{1}{H+\delta H}\der{\psi+\delta \psi}{y} \label{statp}
\eeq 
Using the Taylor development 
\beq 
\fr{1}{H+\delta H}=\fr{1}{H} \biggl[1-\fr{\delta H}{H} +o\biggl(\fr{\norme{\delta H}^2}{\norme{H}^2}\biggr)\biggr] \label{devel}
\eeq
 and neglecting  high order terms, we get from  \rf{statp}
\beqnn
\omega +\delta\omega=\der{}{x} \biggl(\fr{1}{H}-\fr{\delta H}{H^2}\biggr)\der{\psi+\delta\psi}{x} + \der{}{y} \biggl(\fr{1}{H}-\fr{\delta H}{H^2}\biggr)\der{\psi+\delta\psi}{y} 
\eeqnn
The difference between this equation and the equation \rf{stat} is 
\beqr
\delta\omega&=&\der{}{x} \biggl(\fr{1}{H}\biggr)\der{\delta\psi}{x} + \der{}{y} \biggl(\fr{1}{H}\biggr)\der{\delta\psi}{y} -
\der{}{x} \biggl(\fr{\delta H}{H^2}\biggr)\der{\psi}{x} - \der{}{y} \biggl(\fr{\delta H}{H^2}\biggr)\der{\psi}{y}- \nonumber\\
&-&
\der{}{x} \biggl(\fr{\delta H}{H^2}\biggr)\der{\delta\psi}{x} - \der{}{y} \biggl(\fr{\delta H}{H^2}\biggr)\der{\delta\psi}{y} \nonumber
\eeqr
So far, both $\delta\psi$ and $\delta H$ are supposed to be small, we neglect their product and write  briefly
\beq
\nabla \fr{1}{H}\nabla \delta\psi = \delta\omega+ \nabla \fr{\delta H}{H^2}\nabla \psi  \label{pertpsi}
\eeq

This equation allows us to find the perturbation of the streamfunction from perturbations of  vorticity and topography.  

To get the vorticity perturbation, we consider the evolution equation \rf{btp}. As well as above, we write the equation for the perturbed topography using the development \rf{devel} and neglecting high order terms:
\beqr
\der{(\omega+\delta\omega)}{t} &+&\jac(\psi+\delta\psi, \fr{\omega+f}{H}+\fr{\delta\omega}{H}-\fr{\omega+f}{H}\fr{\delta H}{H} )= 
\nonumber\\ &=&
\nu\Delta\omega+\nu\Delta\delta\omega -\sigma\omega-\sigma\delta\omega +curl \frac{\tau }{\rho_0 H_0} \label{perturbed}
\eeqr
The difference between the perturbed equation and the non-perturbed one \rf{btp} writes
\beqnn
\der{\delta\omega}{t} +
\jac(\psi,\fr{\delta\omega}{H}-\fr{\omega+ f_0+\beta y}{H}\fr{\delta H}{H})+
  \jac(\delta\psi, \fr{\omega+ f_0+\beta y}{H})
= \nu\Delta\delta\omega -\sigma\delta\omega+O(\delta^2) 
\eeqnn
in this equation we have also dropped out the term $\jac(\delta\psi,\fr{\delta\omega}{H}-\fr{\omega+ f_0+\beta y}{H}\fr{\delta H}{H})$ that contains the product of functions supposed to be small. 
Using the expression for $\delta\psi $ from \rf{pertpsi}, we get 
\beqr
\der{\delta\omega}{t} &+&
\jac(\psi,
  \fr{\delta\omega}{H}-\fr{\omega+ f_0+\beta y}{H}\fr{\delta H}{H})-
  \jac( \fr{\omega+ f_0+\beta y}{H}, \biggl(\nabla \fr{1}{H}\nabla\biggr)^{-1} \delta\omega ) - \nonumber\\
 &-& \jac(\fr{\omega+ f_0+\beta y}{H},\biggl(\nabla \fr{1}{H}\nabla\biggr)^{-1} \biggl(  \nabla \fr{\delta H}{H^2}\nabla \psi\biggr))
= \nu\Delta\delta\omega-\sigma\delta\omega +O(\delta^2) \nonumber
\eeqr
This equation can be written in a short matricial form
 \beq
 \der{\delta\omega}{t} = A(\psi,\omega)\delta\omega+ 
  B(\psi,\omega)\fr{\delta H}{H} \label{mateq}
 \eeq
 where operators $A$ and $B$ are defined as 
\beqr
A(\psi,\omega) \xi &=& 
-J(\psi,\fr{\xi}{H}) +J( \fr{\omega+ f_0+\beta y}{H}, \biggl(\nabla \fr{1}{H}\nabla\biggr)^{-1} \xi )+\nu\Delta\xi -\sigma\xi\label{A} \\
B(\psi,\omega) \xi&=& 
J(\psi,
  \fr{\omega+ f_0+\beta y}{H}\xi)
  + J(\fr{\omega+ f_0+\beta y}{H},\biggl(\nabla \fr{1}{H}\nabla\biggr)^{-1} \biggl(  \nabla \fr{\xi}{H} \nabla \psi\biggr)) \label{B}
\eeqr
The  system \rf{mateq} starts from the zero initial state $\delta\omega(x,y,0)=0$ because the purpose is confined on the study of the sensitivity of the solution to the topography, rather than to the initial state. We require the perturbed solution $\{\psi+\delta\psi, \omega+\delta\omega\}$ to have the same boundary condition as $\{\psi, \omega\}$ because we do not want to study the model's sensitivity to boundary conditions. Hence,   perturbations $\delta\psi, \delta\omega$ in equations \rf{pertpsi}, \rf{mateq} must satisfy
\beq 
\delta\psi\mid_{\partial\Omega} =0,\;\delta\omega\mid_{\partial\Omega} =0 \label{bcpert}
\eeq 

Analysing the form of the equation \rf{mateq}, we can see the right-hand-side  is  composed by two  terms $A$ and $B$ (\rf{A}, \rf{B}).  The first one, $A$, is responsible for the evolution of a small perturbation by the model's dynamics, while the second one, $B$,  determines the way how the uncertainty is introduced into the model. The first term is similar for any sensitivity analysis, while the second one is specific to the particular variable under study. This term is absent when the sensitivity to    initial point is studied because the uncertainty is introduced only once, at the beginning of the model integration. But, when the uncertainty is presented in the bottom topography or in some other internal  parameter of the model, the perturbation is introduced at each time step and requires an additional operator.

If the reference trajectory is a stationary point, then operators $A$ and $B$ do not depend on time. In this case   the system  \rf{mateq} can be resolved  explicitly. The solution is 
  \beq
 \delta\omega(T)= \underbrace{(e^{TA}-I) A^{-1} B}_{G(T)} \fr{\delta H}{H} = G(T)\fr{\delta H}{H} \label{explicit}
  \eeq

  If the reference trajectory $\{\psi,\omega\}$ is not stationary, discrete time stepping is used to integrate \rf{mateq} using  some numerical scheme. For the simplest Euler scheme, iterations procedure on the part of the reference trajectory $\omega(t_0) ... \omega(t_0+T)$  can be calculated in the following way:
\beqnn  \fr{\delta\omega^{n+1}-\delta\omega^{n}}{\tau} = A(\psi(t_n),\omega(t_n))\delta\omega^n+B(\psi(t_n),\omega(t_n))\fr{\delta H}{H} 
\eeqnn
 Let us suppose that there exists a matrix $G_{n}$ such as  at the $n$th time step 
\beqnn
\delta\omega^{n}= G_{n}\fr{\delta H}{H}
\eeqnn
Then, at the  $n+1$th time step we get
\beqr
 \delta\omega^{n+1} &=& 
 \biggl(I+\tau A(\psi(t_n),\omega(t_n)) \biggr)\delta\omega^{n} + \tau B(\psi(t_n),\omega(t_n))\fr{\delta H}{H} = \nonumber \\
 &=& \biggl[ \biggl(I+\tau A(\psi(t_n),\omega(t_n)) \biggr) G_{n}  + \tau B(\psi(t_n),\omega(t_n))\biggr]\fr{\delta H}{H} = \nonumber \\
 &=&  G_{n+1}\fr{\delta H}{H}   \nonumber
 \eeqr
Thus, the matrix $G_{n+1}$ is calculated as
 \beq
 G_{n+1} = \biggl(I+\tau A(\psi(t_n),\omega(t_n)) \biggr) G_{n}  + \tau B(\psi(t_n),\omega(t_n)) \label{iterations} 
 \eeq
  for  $n = 0,1,.. N=T/\tau$ with  $G_0=0$.   It has to be noted here, the operator $G_{N}$ for a non stationary trajectory depends not only on the time interval $T$, but also on the trajectory part passed by the reference solution $\psi(t),\omega(t), t=t_0\ldots t_0+T$. So far, the solution of the reference system is unique, we can say the operator $G_{N}$  depends  on the initial point $\psi(t_0),\omega(t_0)$. To express this dependence explicitly we shall write the operator $G_{N}$  as $G(t_0,T)$. 
  
  In this paper we are looking for   the most dangerous  perturbation of the topography. That means, the topography's perturbation  of a given  small norm which maximizes the norm of the solution's perturbation $\norme{\delta\omega}$ at the time $T$. 
We can chose any norm in this consideration 
\beq
\norme{\delta\omega}^2 = \spfds{{\cal K} \delta\omega(x,y,t),\delta\omega(x,y,t)}\int_\Omega {\cal K} \omega(x,y,t) \omega(x,y,t)  dx dy
\eeq
with an auto-adjoint  positive definite operator ${\cal K}$. 
Thus, for example,  the energy of the solution is obtained with the inverse Laplace operator ${\cal K}=\Delta^{-1}$, the enstrophy corresponds to identity operator ${\cal K}=I$.    

  In other words we are looking for the 
 \beqr
 \max\fr{\norme{\delta\omega(T)}^2}{\norme{\delta H}^2} &=& 
 \max \fr{\spfds{{\cal K}\delta\omega(T),\delta\omega(T)}}
    {\spfds{{\cal K}\delta H,\delta H}} = \nonumber\\
    &=&
 \max\fr{\spfds{G^*(t_0,T){\cal K}G(t_0,T)\delta H,\delta H}}
    {\spfds{{\cal K}\delta H, \delta H}} =\nonumber\\ 
 &=& 
 \max\fr{\spfds{{\cal K}{\cal K}^{-1}G^*(t_0,T){\cal K}G(t_0,T)\delta H,\delta H}}{\spfds{{\cal K}\delta H, \delta H}} =
 \nonumber\\ 
 &=&
 \max \lambda^2({\cal K}^{-1}G^*(t_0,T){\cal K}G(t_0,T))\label{maxlam}
 \eeqr
 where $\lambda^2(G^*(t_0,T)G(t_0,T))$ are defined as eigenvalues of the problem 
 \beqnn {\cal K}^{-1}G^*(t_0,T){\cal K}G(t_0,T)\varphi_i = \lambda_i^2 \varphi_i  
 \eeqnn
Below the enstrophy of the solution  is considered as its norm 
\beq
\norme{\delta\omega}^2 = \int_\Omega \omega^2(x,y,t) dx dy
\eeq
and the corresponding eigenvalues problem 
 \beq G^*(t_0,T)G(t_0,T)\varphi_i = \lambda_i^2 \varphi_i  \label{egvlpb}
 \eeq
 
The eigenvalues show the growth rate of different modes of  the perturbation. 
Taking into account the expression for $G(t_0,T)$ in the stationary case \rf{explicit} we see that in the infinite time limit $ T \longrightarrow \infty$, the operator tends to $ B^*(A^*)^{-1}(e^{TA}-I)^*(e^{TA}-I) A^{-1} B$. The only dependence on $T$ is associated with the exponential. If we consider the expression $ lim_{T\longrightarrow \infty}\fr{\ln\lambda}{T}$ we see the  value of this expression tends to the $ lim_{T\longrightarrow \infty}\fr{\ln\nu}{T}$, where $\nu$ are eigenvalues of the pure exponential operator $e^{TA}$:
\beqnn (e^{TA})^*e^{TA}\varphi_i = \nu_i^2 \varphi_i
 \eeqnn
 That means the growth rate of the perturbation is determined by the maximal Lyapunov exponent of the model, i.e. it is the same rate as in the study of sensitivity to initial state. This reasonable conclusion shows the system develops its own instability. No matter what was  the source of  perturbation,  it will behave according to the system's internal instability modes on  long time scales.  
 
 The same behavior can be expected in the non stationary case. Despite there is no explicit exponent in the expression \rf{iterations}, it is well known the   nonlinear model on a strange attractor reveals exponential  growth rate of perturbations on infinite time scales. So, the intrinsic instability of the non stationary model will also dominate on long times. 
  
  However, on short time scales the instability will show different behavior. This is natural because the matrices $A^{-1}$ and $ B$ in \rf{explicit} are comparable with the exponent on  short time scales, i.e.  the source of the perturbation is important on these scales.

\section{Time dependent solution}

The model has been discretised in space using finite elements method. Details of  discretisation and construction of matrices $A$ and $B$ (\rf{A}, \rf{B}) are shown in the Appendix. 

The grids used in this paper are  presented in \rfg{fig1}. 
The triangulation of the square is composed of 206 triangles.  The integration points set, being a union of vertices and mi-edges of triangles, 
counts 445 nodes. The resolution of the grid varies between $1/80$ of the side length (about 50 km for the square of 4000 km) near the western boundary and $1/10$ of the side length (about 400 km) near the eastern one.

The triangulation of the North Atlantic is  also performed with a grid refinement near the American coast. This triangulation is composed of 195 triangles and 436 points. The resolution of this grid is about the same, i.e. about 40 km near the American coast and about 400 km near the European one.

\figureleft{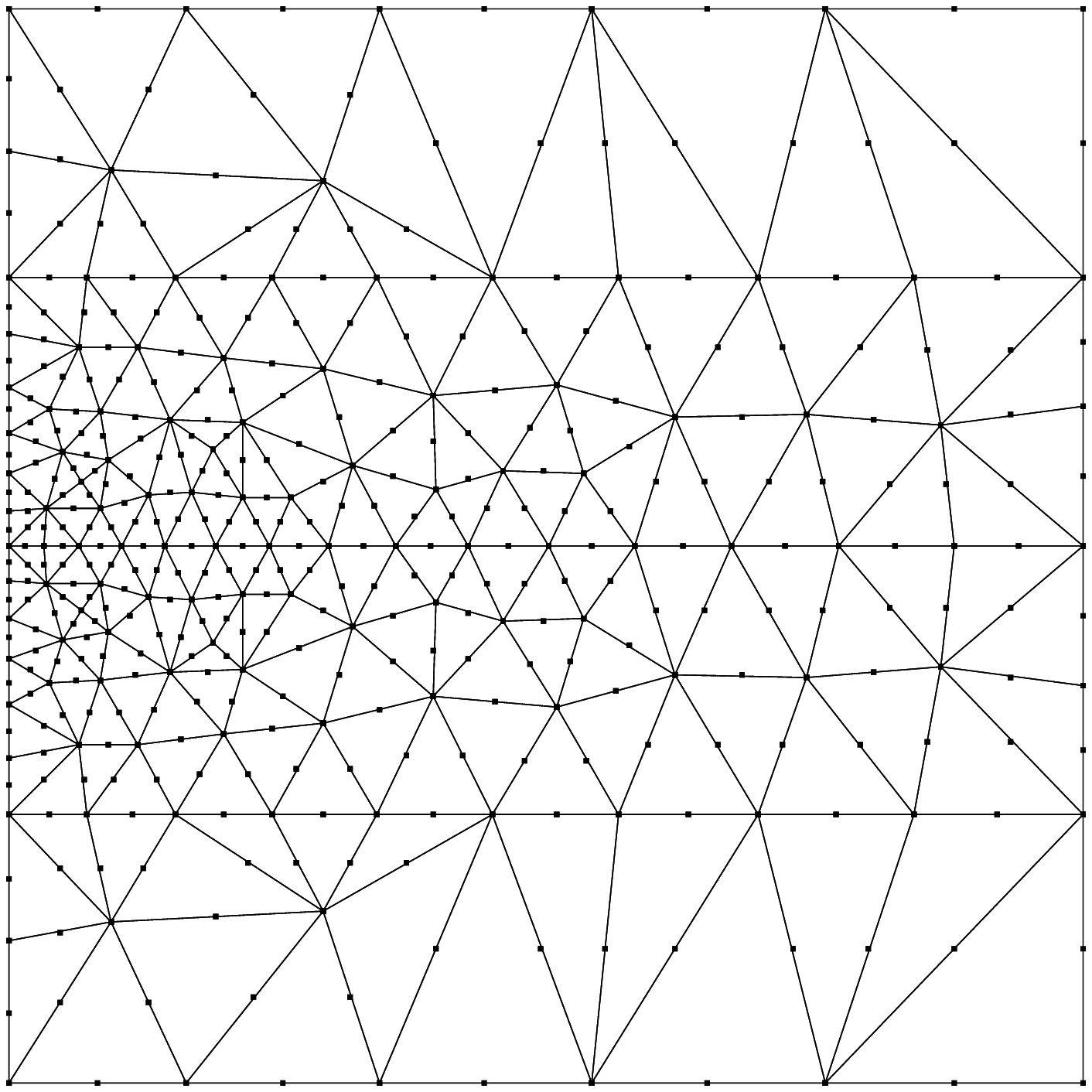}{Triangulation of the square.}{fig1}  
\figureright{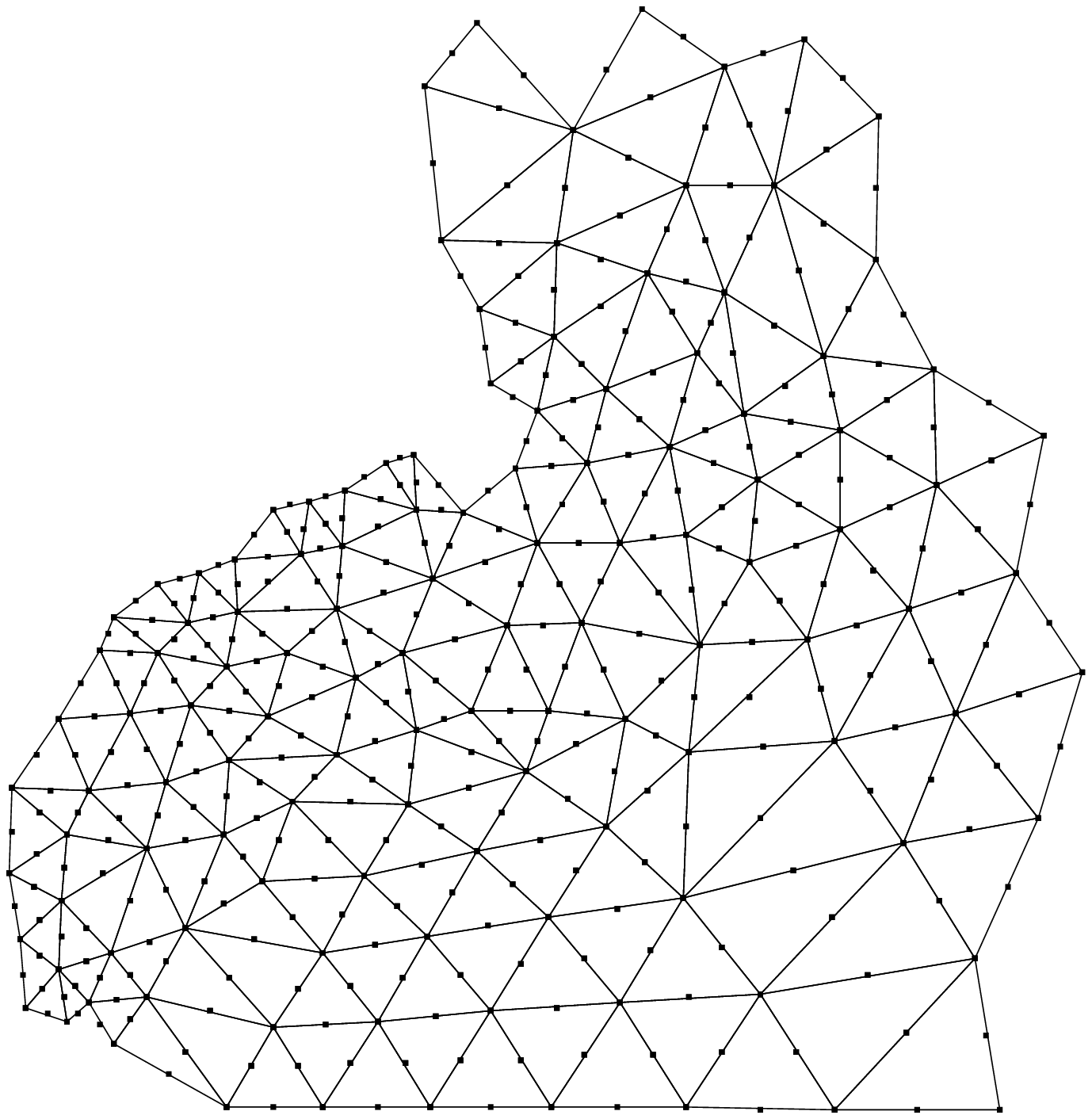}{ Triangulation of the North Atlantic region.}

In the experiment with the square box we take the characteristic length of the basin  $L=  4000$ km.  The bottom is supposed to be either flat or with regular sinusoidal topography.  
We take a steady zonal wind as forcing with  now classical two
gyre antisymmetric pattern.  This is seen as a schematic 
pattern for the mean curl of the wind stress over the North 
Atlantic ocean in middle latitudes. Its magnitude is equal
to 
 \beq 
{\cal F}(x,y) =-\fr{2\pi\tau_0}{L} \sin \fr{2\pi y}{L} \label{forc1}
\eeq 
where $\tau_0=1.1\fr{dyne}{cm^2}$ is the characteristic wind tension on the surface.

	The coefficient of Eckman dissipation we chose as
$\sigma=5\times10^{-8}s^{-1}$, which corresponds to the damping time-scale 
$ T_\sigma = 2\times10^7s \sim 200\mbox{ days}.$ The lateral friction coefficient $\nu$ has been chosen in order to avoid numerical instability which occurs due to the concentration of   variability of the model at grid scales. This value has been taken to be $\nu=500\fr{m^2}{s}$, that corresponds to the damping time scale $T_{\nu}= 3$ days for a  wave of 100 km length. 

Despite this simplified geometry
accompanied by now classical test forcing ~\rf{forc1}, this ``academic" case has been intensively investigated over the last 15 years in order to study the role of mesoscale eddies in the ocean circulation. This configuration helps us to see the sensitivity in a very well described case.

The second experiment is carried out in a more realistic configuration. The domain was chosen to   approximate the North Atlantic region. We assume that the domain is  comprised  in the rectangular between $78^0W \ldots 3^0W$ in longitude and
$15^0N \ldots 65^0N$ in latitude. The boundary of the basin corresponds to the 1 km depth isobath of the ocean. 

To obtain the  forcing in this experiment we have used the data set ``Monthly Global Ocean Wind Stress Components" prepared and maintained by the Data Support Section, Scientific Computing Division, National Center for Atmospheric Research. These data have been prepared by the routine described in \cite{hellerman}. From this data set we choose the mean January  wind stress components $\tau_x$ and $\tau_y$ over the North Atlantic based on 1870-1976 surface observations. These data are presented on the $2^0\times 2^0$ grid. 

The forcing  in this experiment is calculated from these data as
\beqr
{\cal F}(x,y) = -\fr{\partial \tau_x}{\partial \varphi} +
 \fr{1}{\cos\varphi} \fr{\partial \tau_y}{\partial \lambda}, \\
\varphi=20^0+y\times 50^0/L,  \lambda= -40^0+\fr{x\times 50^0/L}{\cos\varphi}
\nonumber \eeqr
where $L=5500$km is a characteristic length of the basin. The spatial configuration of the forcing is presented in \rfg{topo}B.

The bottom topography has been  interpolated from the   ETOPO5 5-minute gridded elevation data \cite{etopo5}.  It  is shown in \rfg{topo}A. 

We chose the coefficient of Eckman dissipation  to be the same as in the previous experiment. 
$\sigma=5\times10^{-8}s^{-1}.$ The lateral friction coefficient $A$ has been chosen in order to avoid numerical instability which occurs due to the concentration of   variability of the model at grid scales. This value has been taken to be $A=300\fr{m^2}{s}$, that corresponds to the damping time scale $T_A= 6$ days for a  wave of 100 km length.

\figureleft{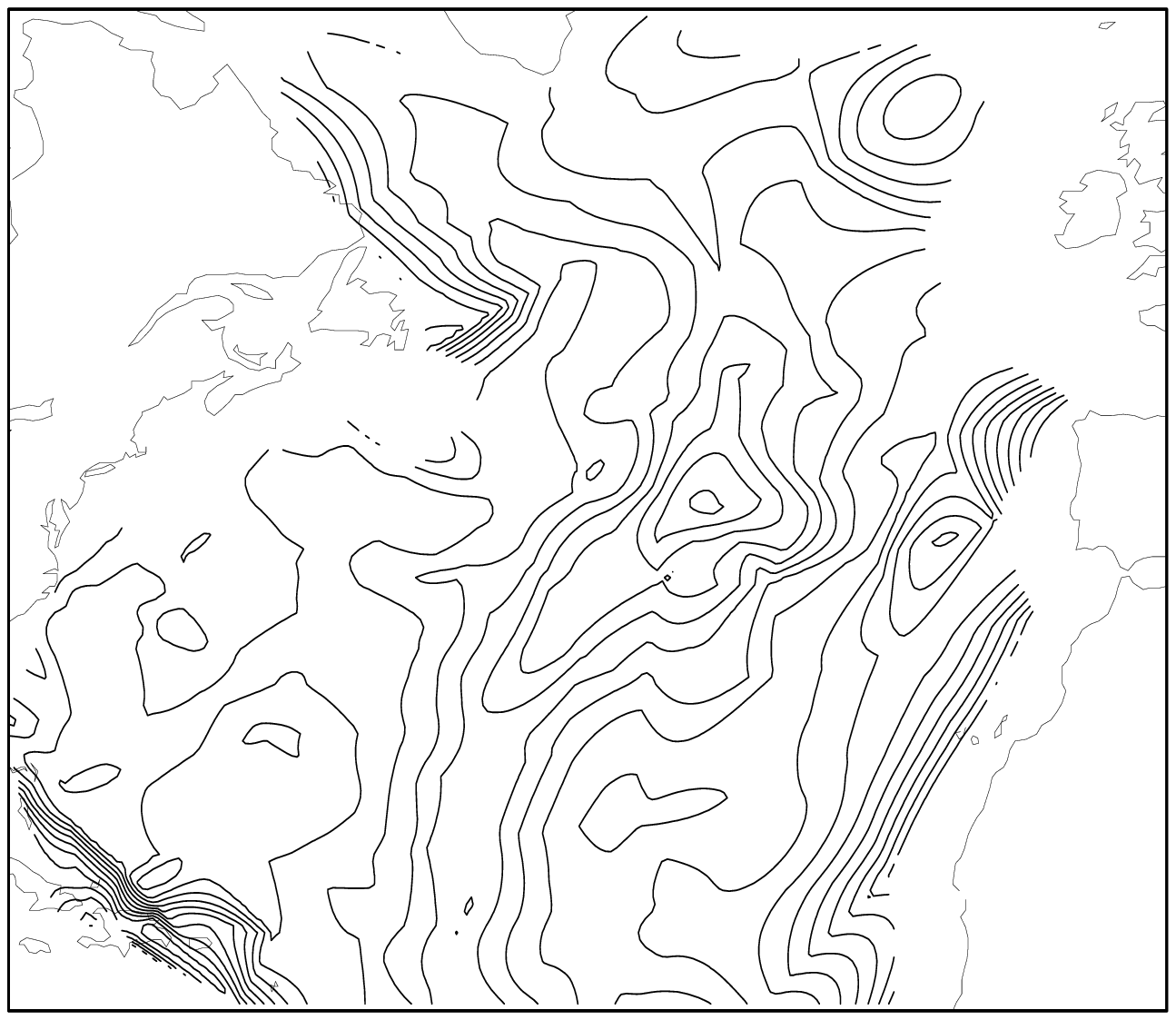}{Bottom topography. Contours from  1000 to 6000m, contour interval 500 m. }{topo}  
\figureright{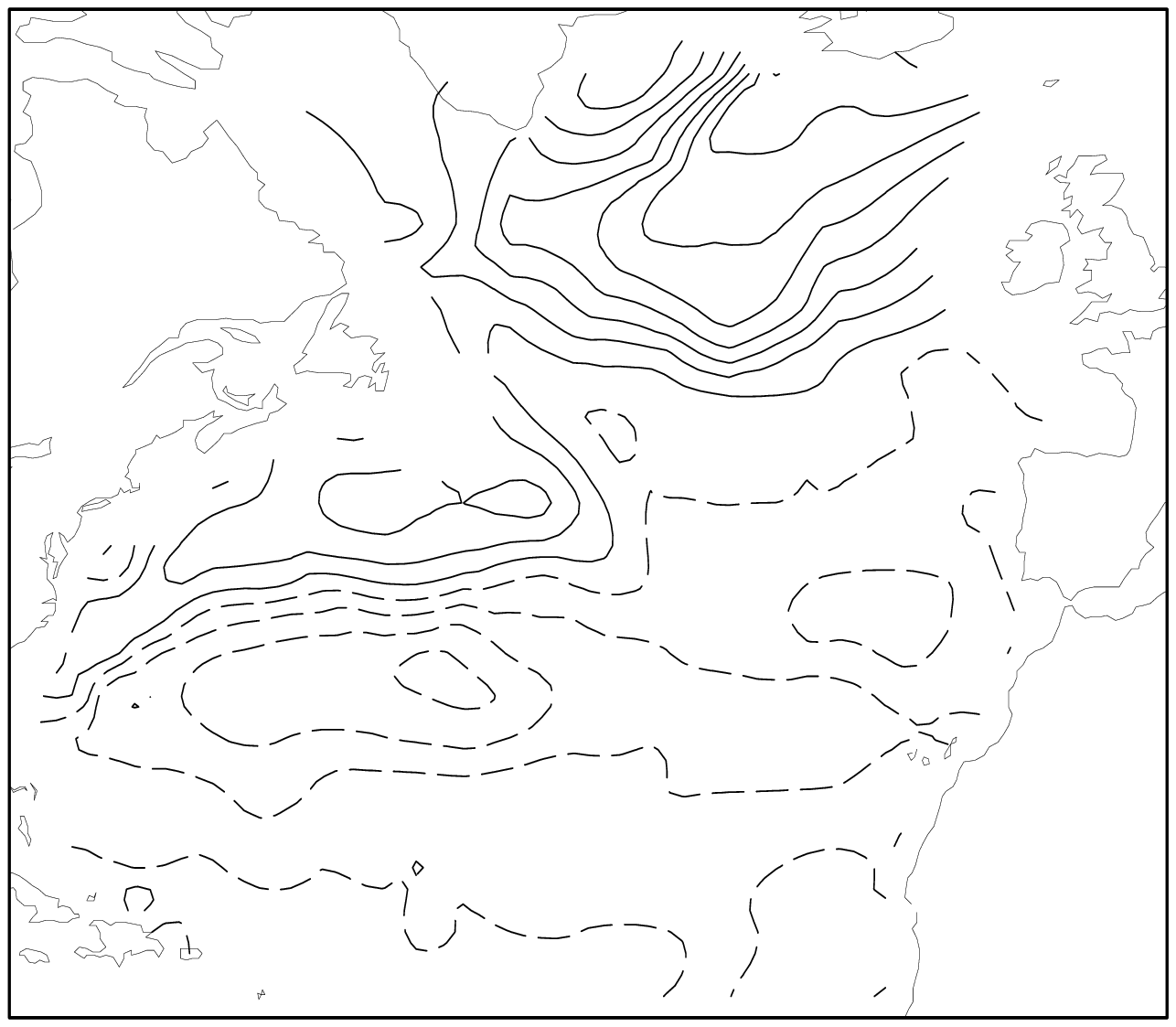}{ Forcing of the model. Contours from -1.5 to 2.4 $\fr{dyne}{cm^2}$ interval 0.3.} 

We  intend to perform several experiments to see the sensitivity  of the model's solution to the bottom topography. First, we study the dependence of the largest singular values of the matrix $G(t_0,T)$ \rf{maxlam} on the initial point of the reference trajectory $t_0$. 

To obtain the initial point in  both experiments, the model has been integrated during 20 years from the zero state. We suppose that after this period the spin-up phase is over and the solution of the model reaches its attractor. After the spin-up, the model is integrated during 204.8 days with the time step 0.1 day. This part of trajectory composed with 2048 samples is used as $\psi(x,y,t)$ in \rf{mateq} to construct  the matrix $G(t_0,T)$ following the procedure \rf{iterations}.  Average streamfunction's  plots for the square box and for the North Atlantic region are  shown in \rfg{psisq}A, \rfg{psiatl}A respectively. To see the length of the part of trajectory under consideration, we plot the kinetic energy versus the enstrophy of the solution in \rfg{psisq}B, \rfg{psiatl}B. 
\beqr
E_k(t)&=&\int_\Omega \psi(x,y,t)\omega(x,y,t) dxdy = \int_\Omega H(x,y)(u^2(x,y,t)+v^2(x,y,t)) dxdy  \\
\eta(t)&=&\int_\Omega \omega^2(x,y,t) dxdy 
\eeqr

\figureleft{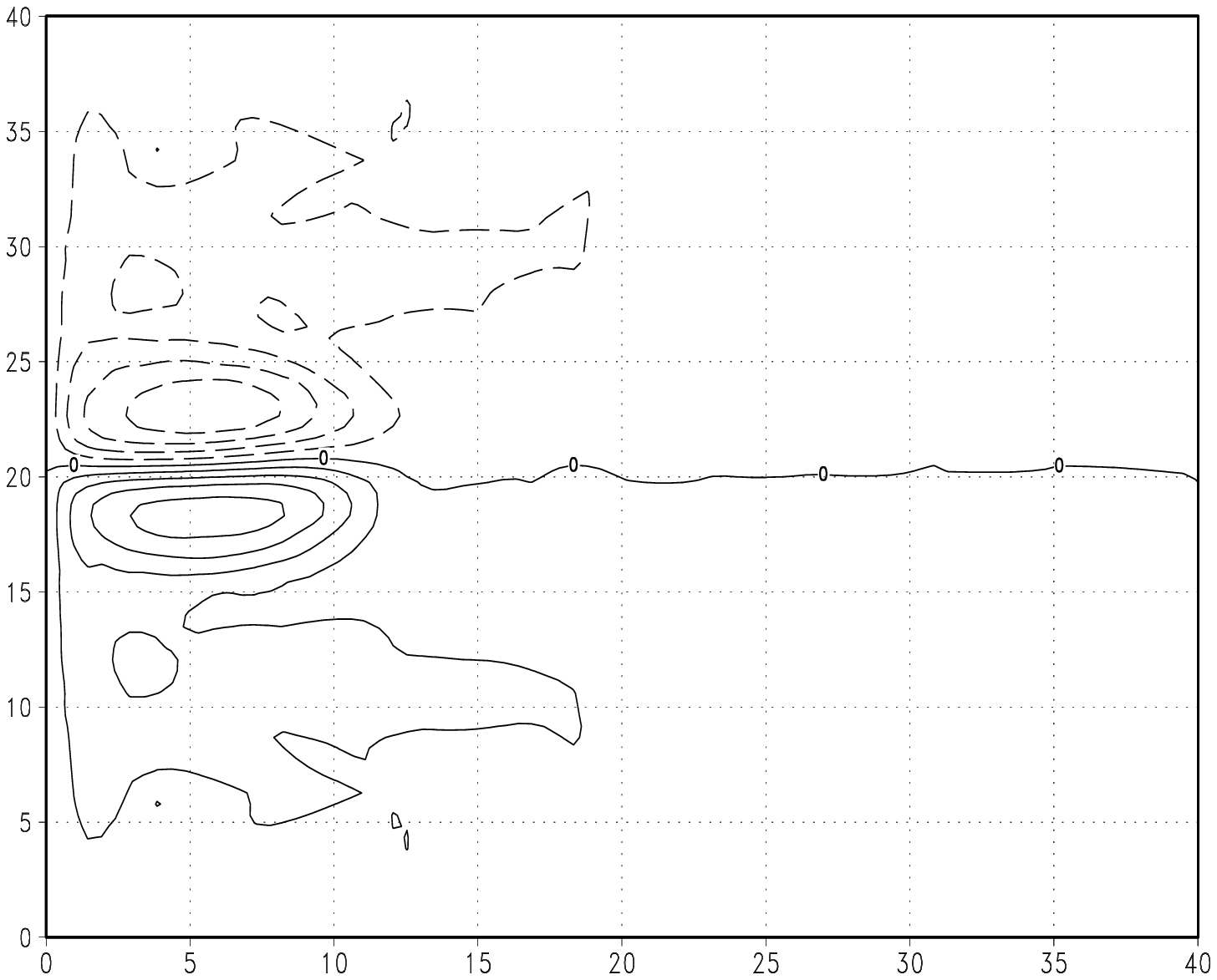}{Mean streamfunction in the square box. 
   Contours from $-8\tm 10^8$ to $8\tm 10^8$. 
   Contour's interval $2\tm 10^8$}{psisq}  
\figureright{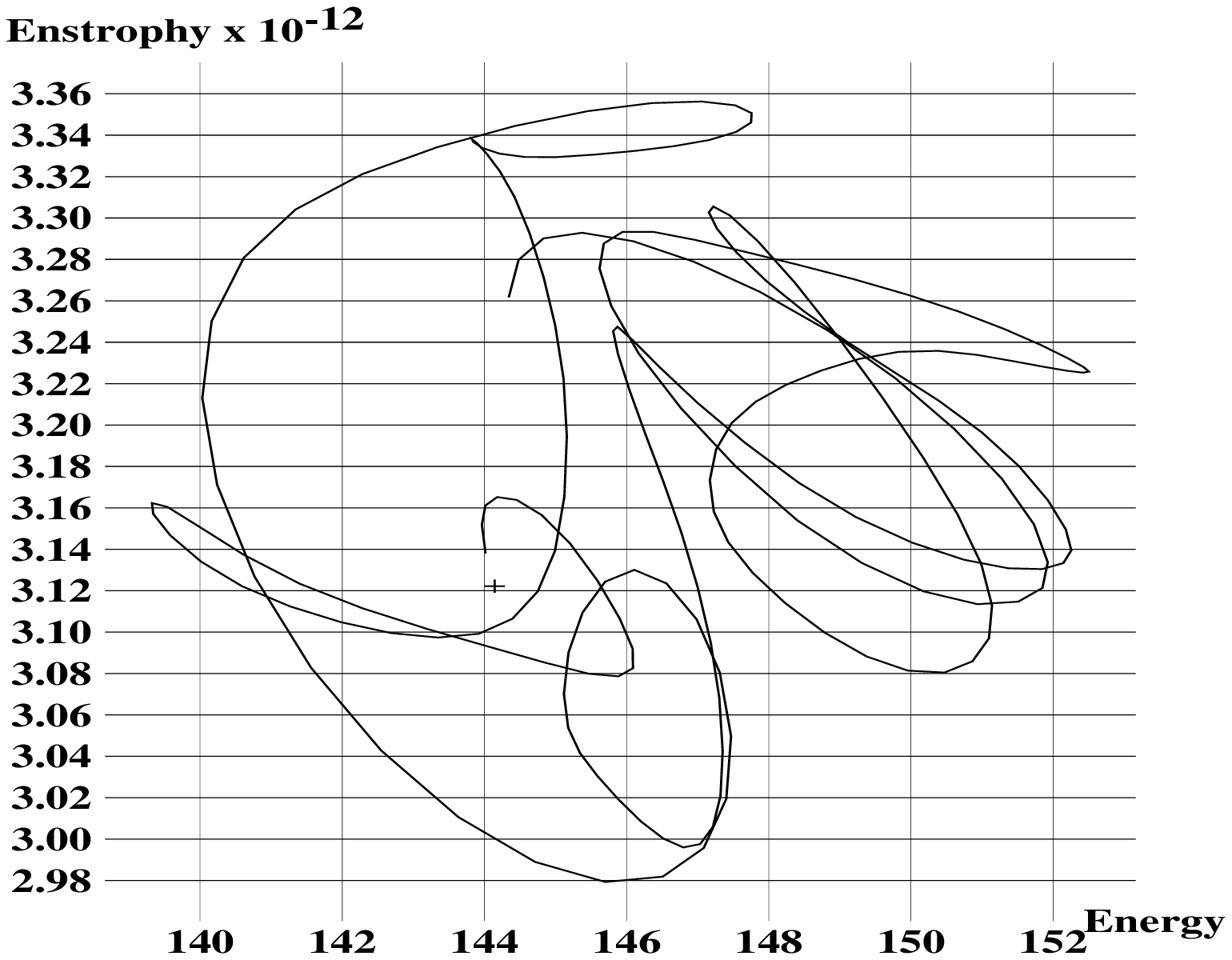}{ Enstrophy-energy plot for 204.8 days trajectory in the square box.}

\figureleft{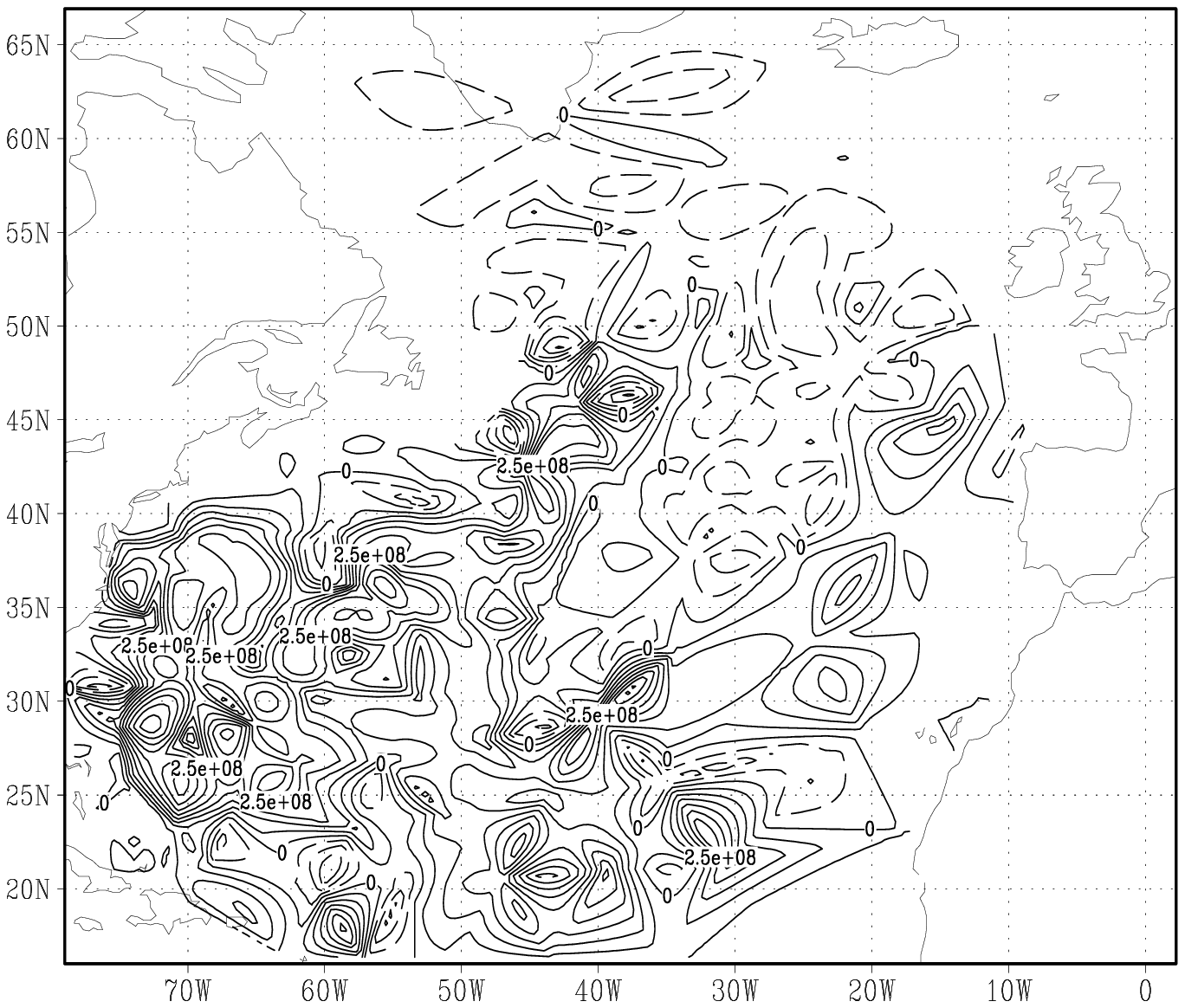}{Mean streamfunction in the North Atlantic.}{psiatl}  
\figureright{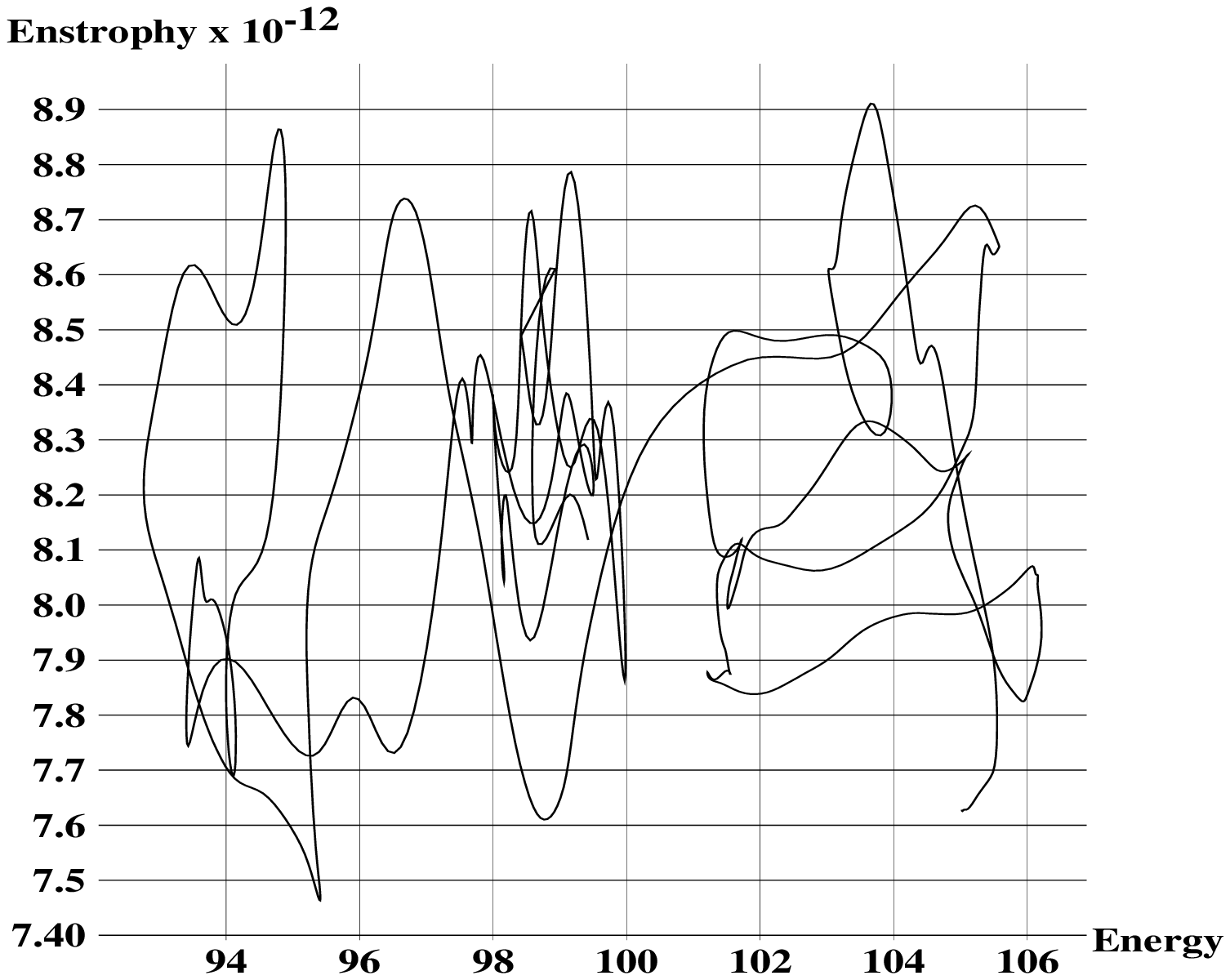}{ Enstrophy-energy plot for 204.8 days trajectory in the North Atlantic.}

 In this sequence of 2048 samples we choose the   set of initial points $t_0$ spaced by $T$. Each interval between two adjacent points is used to construct the matrix $G(t_0,T)$.    Experiments with 2 values of $T$ were carried out: $T=0.8$ and $ 12.8$ days, i.e. the whole sequence was divided respectively to 256 and 16 subintervals of 8 and 128 points each. The value of $T$ we shall call the error growing time, because it is during this time the perturbation $\delta\omega$ is allowed to grow.  Evolution  of the 5 largest singular values for each subinterval are shown in \rfg{evevolsq}, \rfg{evevolatl} as functions of the number of the subinterval in the sequence, or in other words,  of the reference model's time. 

\figureleft{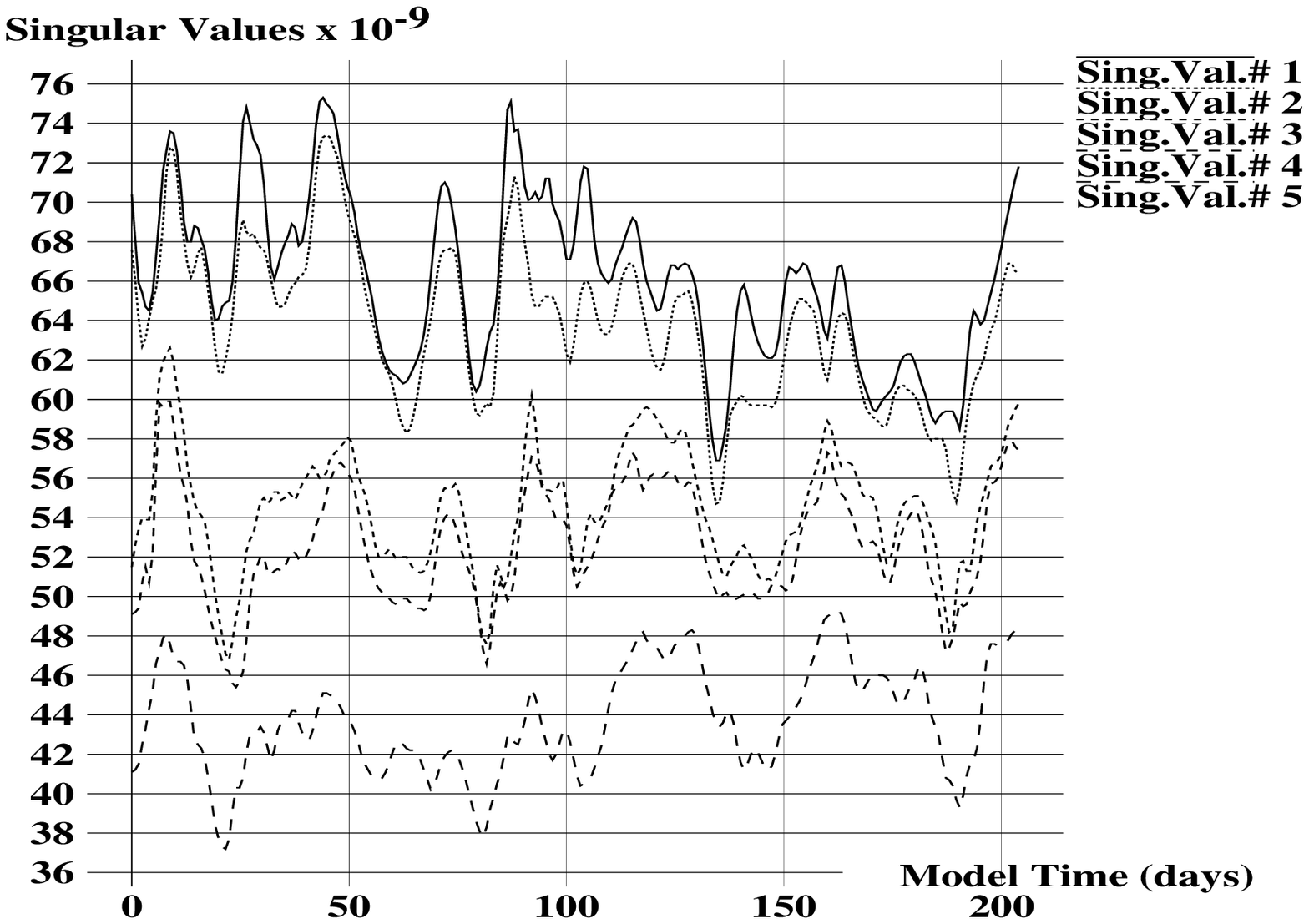}{Evolution of the 5 largest singular values of the matrix $G(t_0,0.8)$ for $0<t_0<204.8$ days in the square box.}{evevolsq}  
\figureright{fig5b.eps}{ Evolution of the 5 largest singular values of the matrix $G(t_0,12.8)$ for $0<t_0<204.8$ days in the square box.}

\figureleft{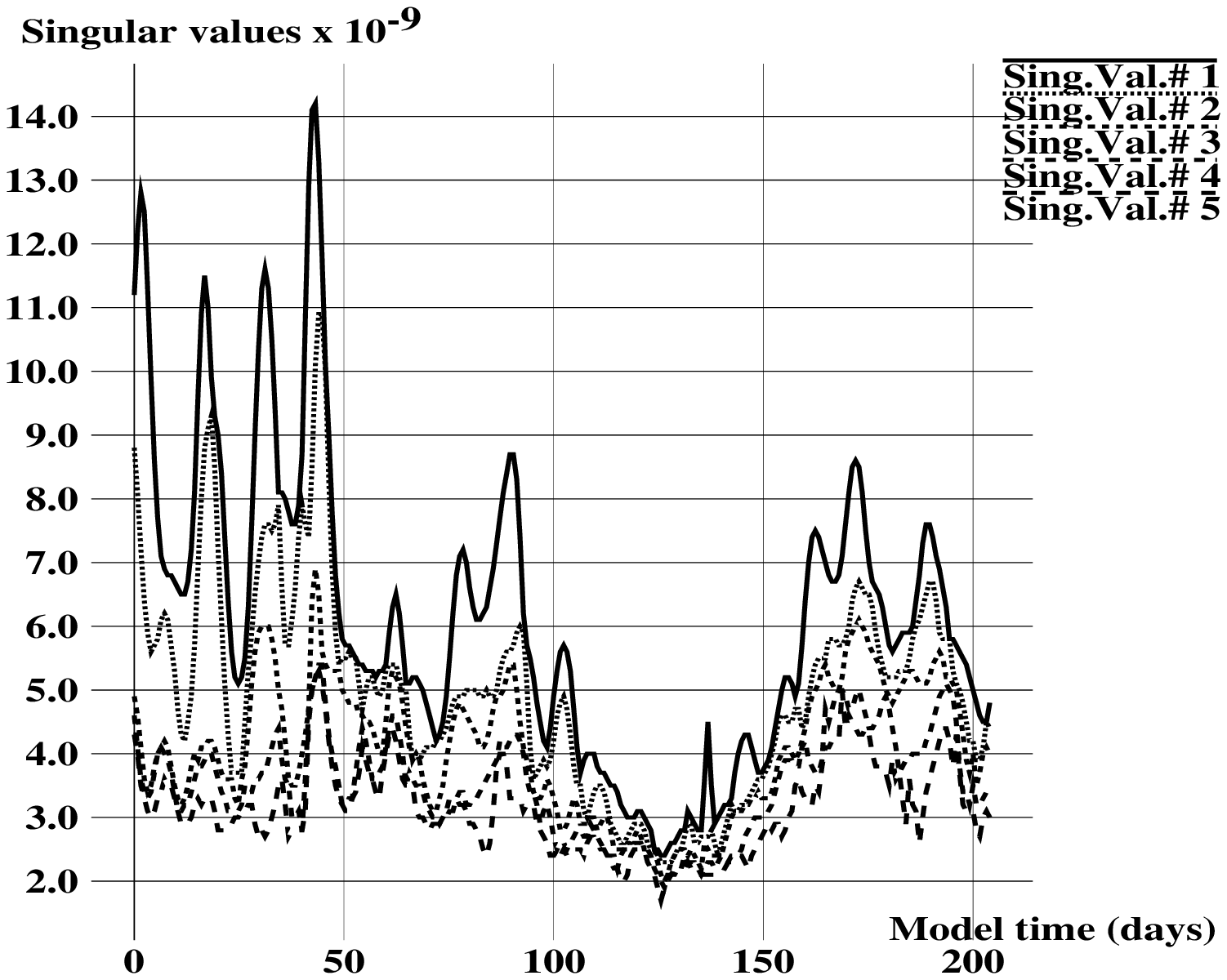}{Evolution of the 5 largest eigenvalues of the matrix $G(t_0,0.8)$ for $0<t_0<204.8$ days in the North Atlantic.}{evevolatl}  
\figureright{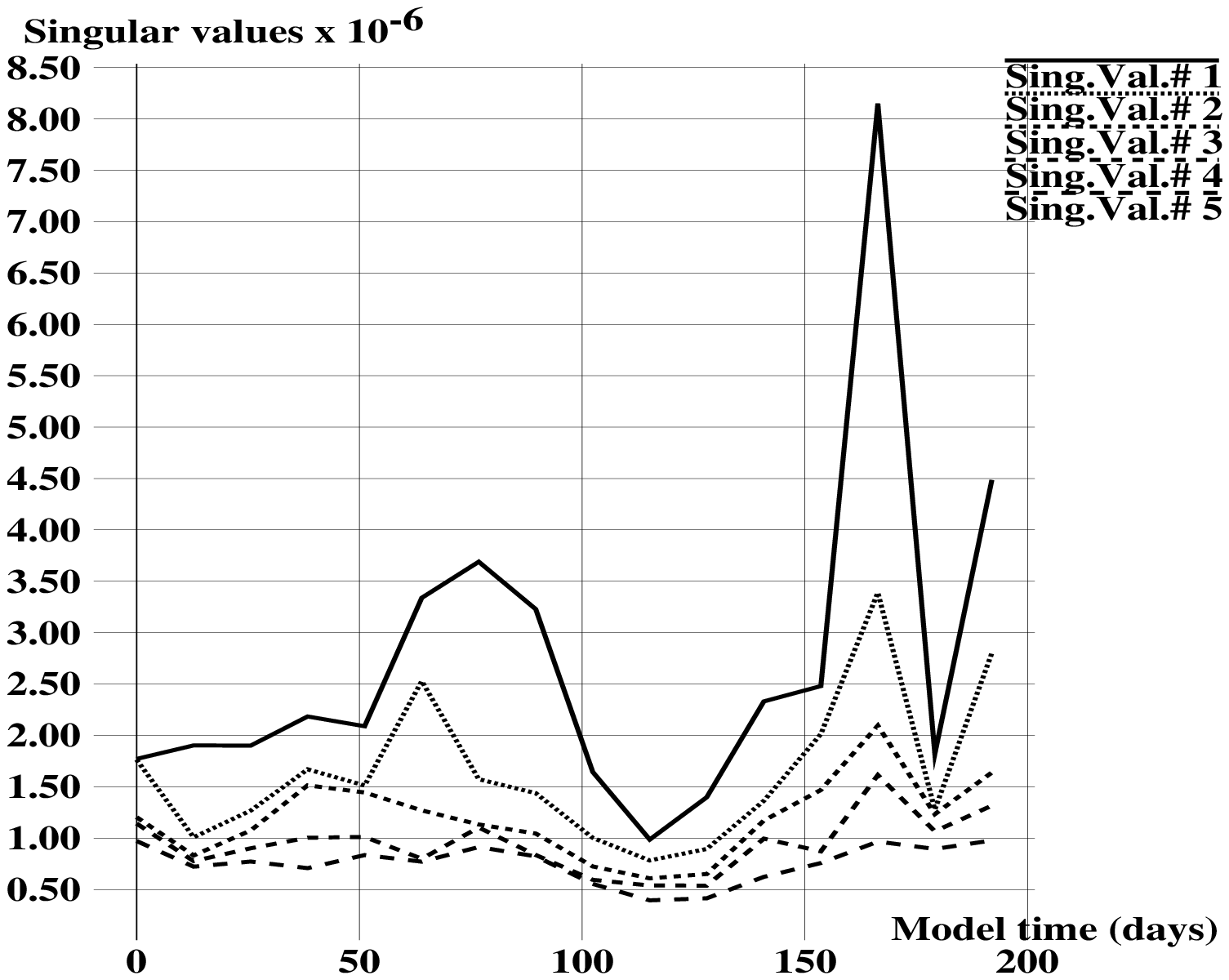}{ Evolution of the 5 largest eigenvalues of the matrix $G(t_0,12.8)$ for $0<t_0<204.8$ days in the North Atlantic.} 

We see the sensitivity of the model's solution to the topography is almost uniform in the square box. The maximum  of the first singular value is approximately only twice its minimum. This is not the case in the experiment with the North Atlantic, where variations have much larger amplitude. We can see in \rfg{evevolatl} there exists a situation with very low sensitivity to the bottom topography. Approximately at 125th day  all largest  singular values in  experiments for both $T=0.8$ and $T=12.8$ days  have a clear minimum. Comparing  the most and the least sensitive situations during the same run, we see the maximum of the first singular value is 7-8 times higher than its minimum. 

Together with  largest singular values of the operator $G(t_0,T)$ \rf{maxlam}, we analyze  corresponding singular vectors. These vectors represent the most sensitive modes of the solution. We calculate these vectors  for both error growing times $T=0.8$ and $12.8$ days. 

An example of the most sensitive singular mode for $T=0.8$ days and for $T=12.8$ days in the square box is
shown in \rfg{efunsq}. Only a  part of the total region, corresponding to the jet-stream near the eastern boundary is presented in these figures.  One can see that for short error growing time (i.e. small values of $T$) the mode is concentrated at just  several points on the grid.  However, for long error growing time of twelve days, the eigenmode occupies more important region. But, in both cases,  the eigenmode is concentrated  near the jet-stream part of the domain.

\figureleft{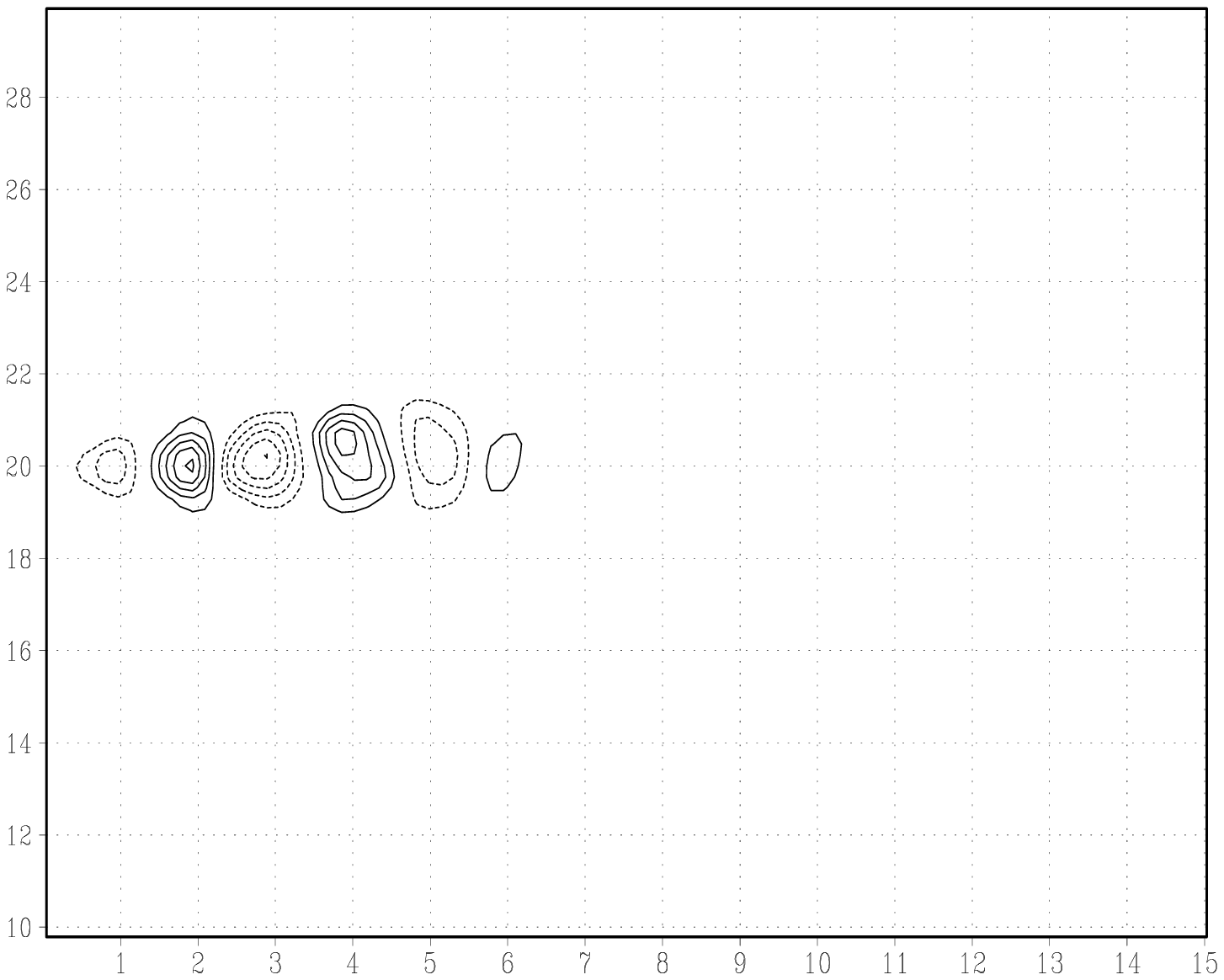}{The most sensitive singular mode for the error growing time $T=0.8$ days. Singular value $7.8\tm 10^{-8}$}{efunsq}  
\figureright{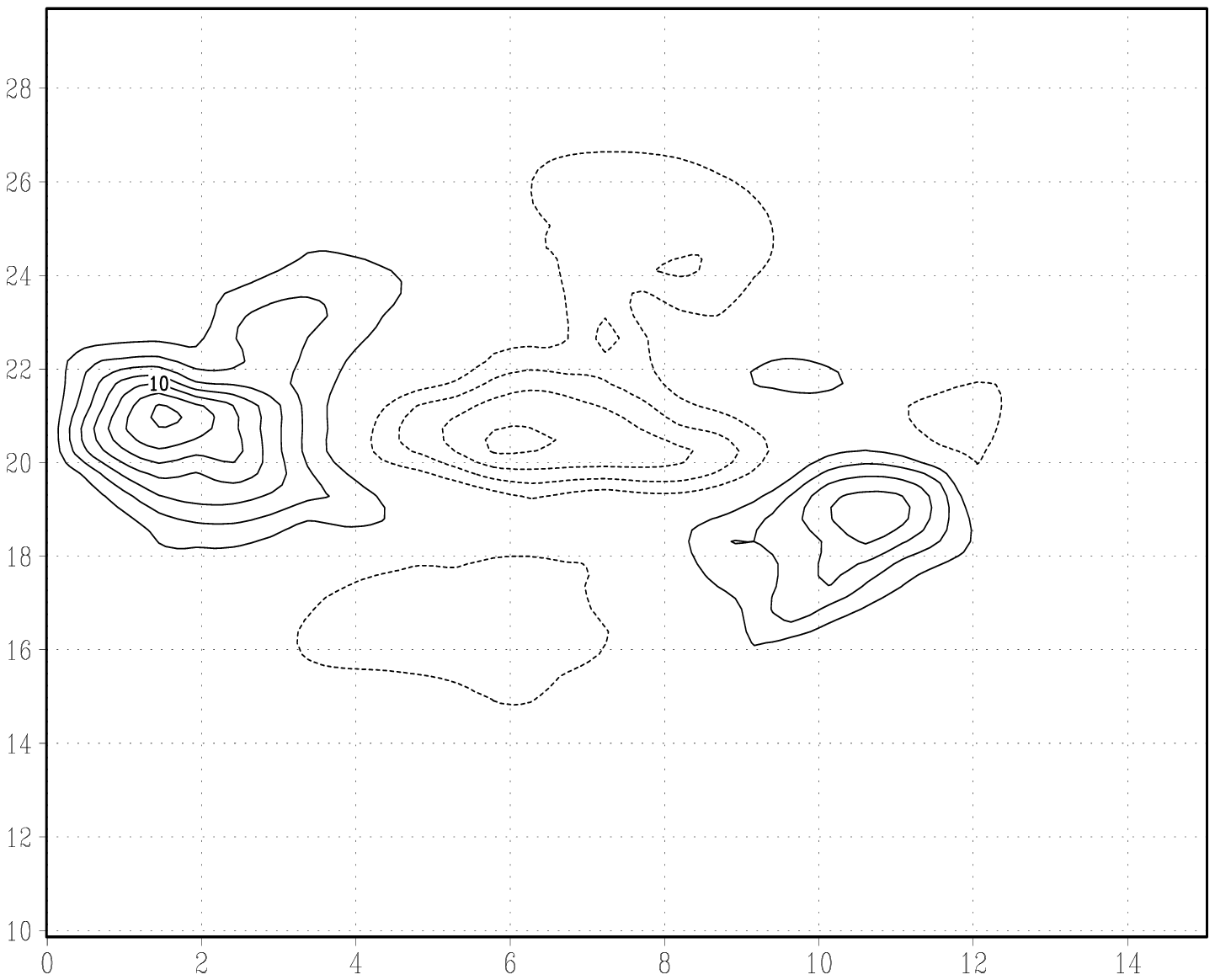}{The most sensitive singular mode for the error growing time $T=12.8$ days. Singular value $9.2\tm 10^{-6}$}

On the other hand, there exists also insensitive modes. Their corresponding singular values are equal to 0 and any perturbation of the model's topography by one of these modes has no impact on the flow. These modes form the kernel of the operator. 

One of these modes can be easily seen from a simple analysis of the model \rf{sw}. If we add to the topography $H$ some perturbation which is proportional to $H$ itself $\delta H =\alpha H$, the model remains the same. Only the third equation is multiplied by $1+\alpha$ in this case and that does not disturb the equality to 0. Hence, the model exhibits no sensitivity to the perturbation $\delta H =\alpha H$ and this mode belongs to the kernel of the operator $G(t_0,T)$ \rf{iterations}. 

Another set of insensitive modes results from the boundary conditions imposed on the vorticity $\omega$ in the equation \rf{btp}. As topography $H$ and it's perturbation $\delta H$ are defined in closed the  domain $\Omega$ including its boundary. But the boundary conditions on $\omega$  require  that $\delta \omega$ is equal to zero on the boundary because both original equation \rf{btp} and perturbed one \rf{perturbed} must follow the same boundary conditions.  The discretised operator $G(t_0,T)$ is represented, hence, by a rectangular matrix with $N_0$ strings and $N$ columns, where $N_0$ is the number of internal points of the domain $\Omega$ and  $N$ is the total number of discretisation points including boundary. The operator $G^*(t_0,T)G(t_0,T)$ \rf{egvlpb} is a square $N\tm N$ operator possessing as many zero eigenvalues as $N-N_0$. These kernel's modes are concentrated on the boundary of the domain.

Considering  internal part of the domain, we can also see the difference in sensitivity.  Least sensitive modes, corresponding to smallest non null singular values represent  perturbations of topography to which the model is very little sensitive. 

This fact is illustrated in \rfg{evtsp} where all non null eigenvalues of the operator $G^*(T)G(T)$ are shown. 
The beginning of the spectrum is shown on the zoom. We can see  singular values decrease rapidly for $T=12.8$ days. The 8th singular value is already 10 times lower than the first one in the square box. In the North Atlantic there exists an outstanding very sensitive first mode which singular value is 10 times  the second one.  So, we need just several  modes to  approximate the whole error behavior. For shorter error growing times (like $T=0.8$ days) this initial decrease is  less sharp and the sensitivity is more uniform. 

In the middle of the spectrum we find a slowly decreasing sequence of singular values. This part represents singular modes with low sensitivity.  And at the end of the spectrum we see the sharp decrease and  several modes with outstandingly low sensitivity.

\figureleft{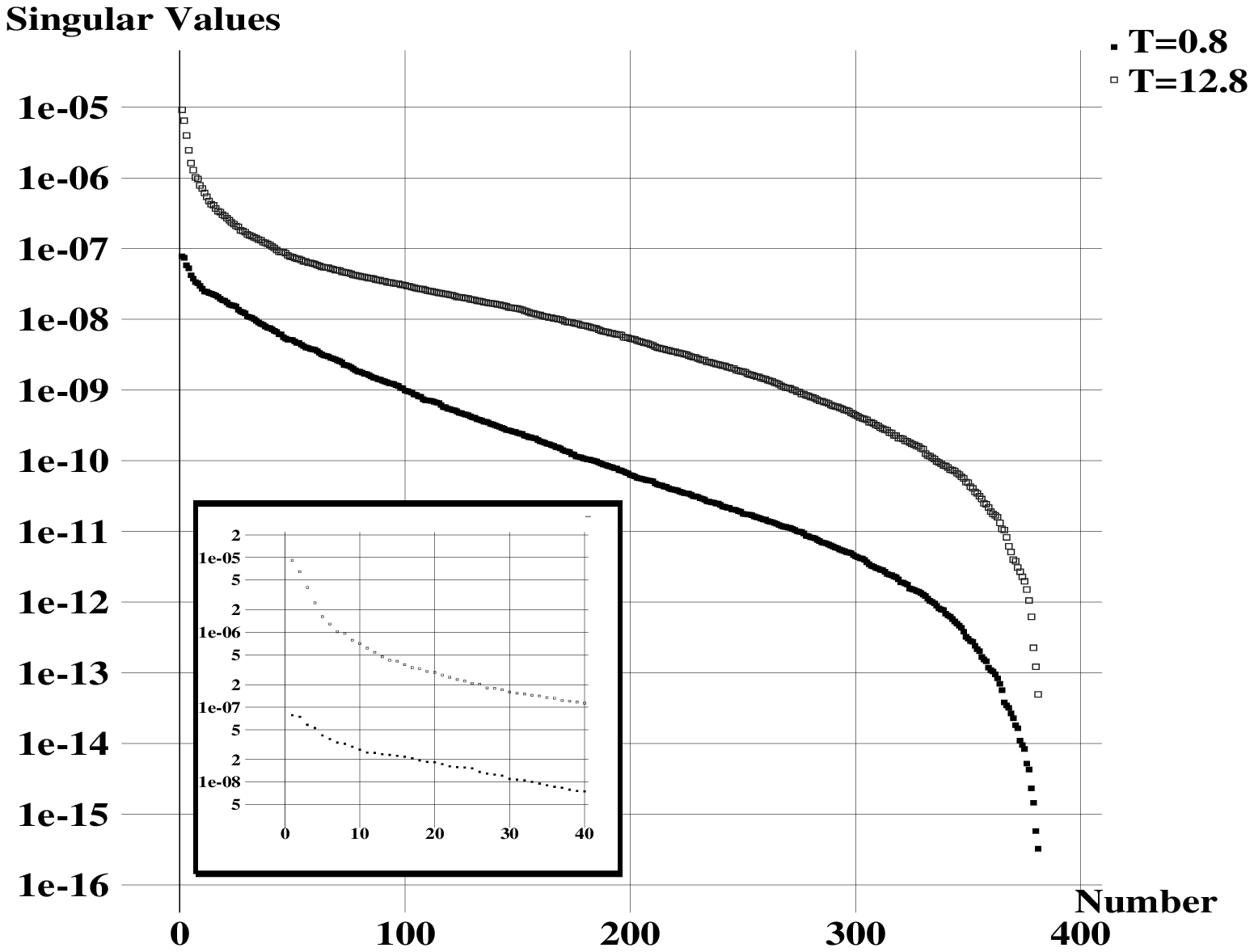}{Singular values of the matrix $G(T)$   for $T=0.1$ and $T=12.8$ days in the square box.}{evtsp}  
\figureright{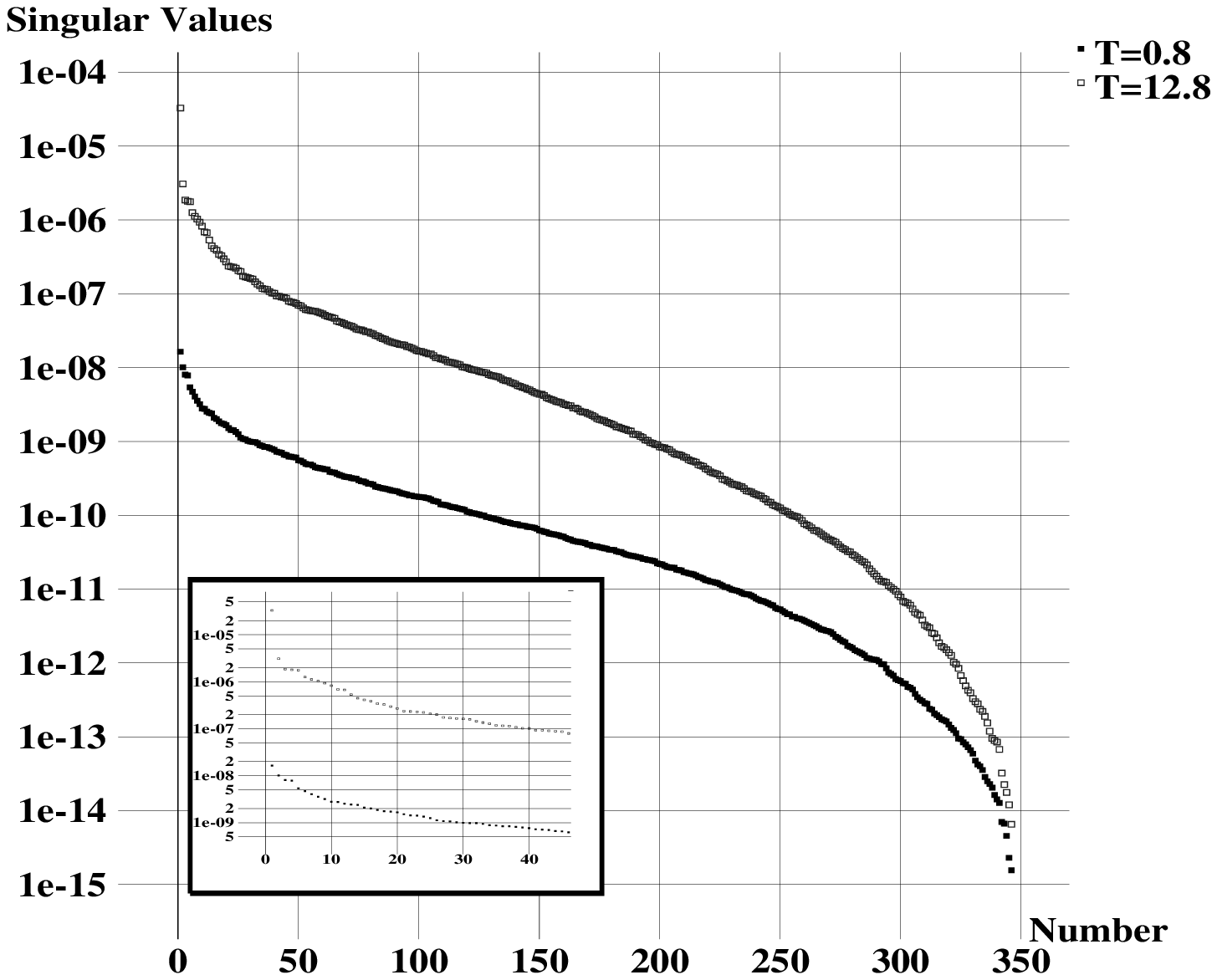}{ Singular values of the matrix $G(T)$  for $T=0.1$ and $T=12.8$ days in the North Atlantic.}

These modes differ from modes in the kernel of the operator. They have no intuitive explanation and they are not concentrated on the boundary. In fact, they  are not localized in space.  They occupy almost the whole domain, especially regions where the flow is smooth and laminar. However, these patterns can not be considered as a grid noise. Their patterns  for both $T=0.8$ and $ 12.8$ days are shown in \rfg{efunfsq}. Contrary to \rfg{efunsq}, whole domain is plotted in these figures.

\figureleft{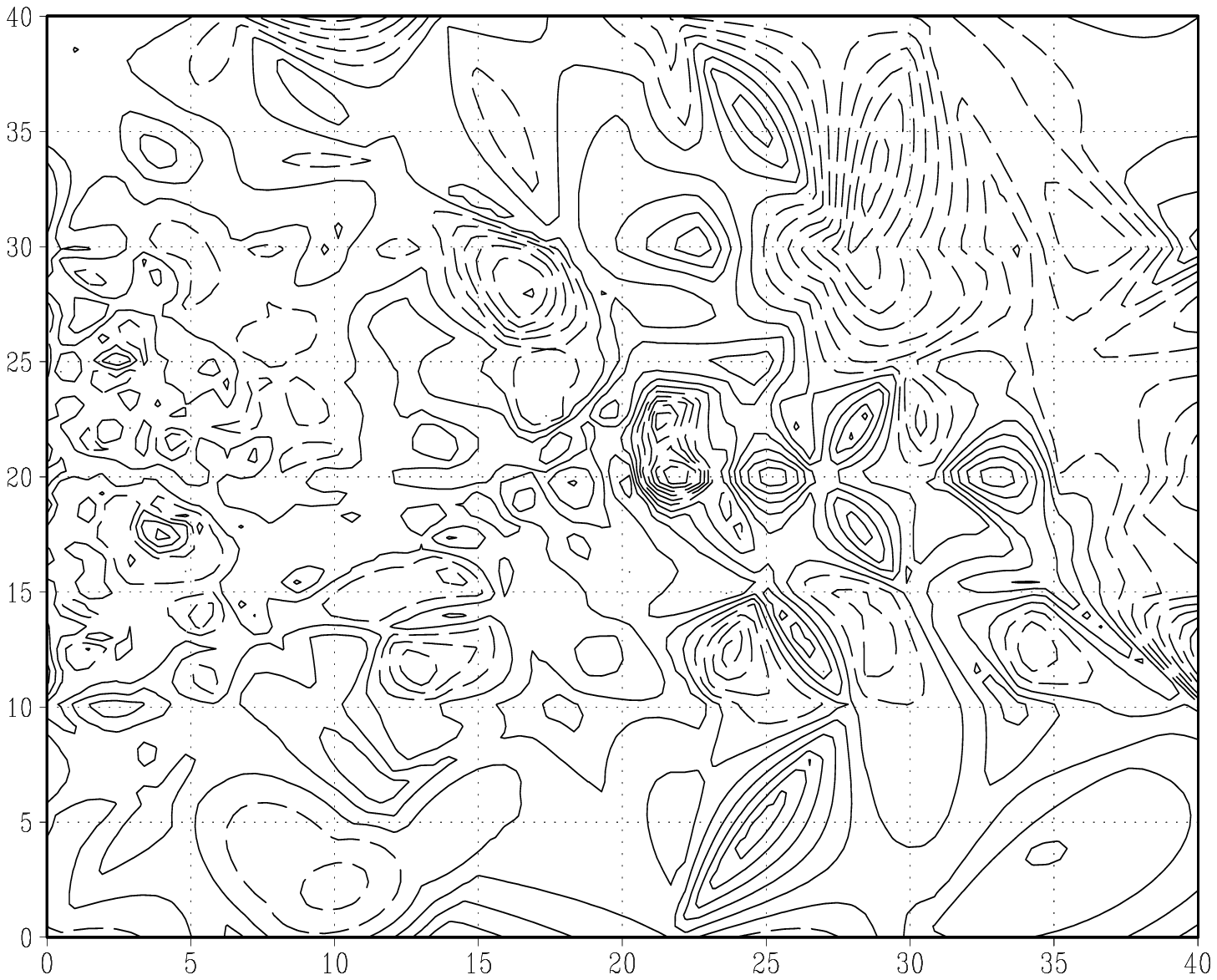}{The most insensitive singular mode for the error growing time $T=0.8$ days. Singular value $3.2\tm 10^{-16}$}{efunfsq}  
\figureright{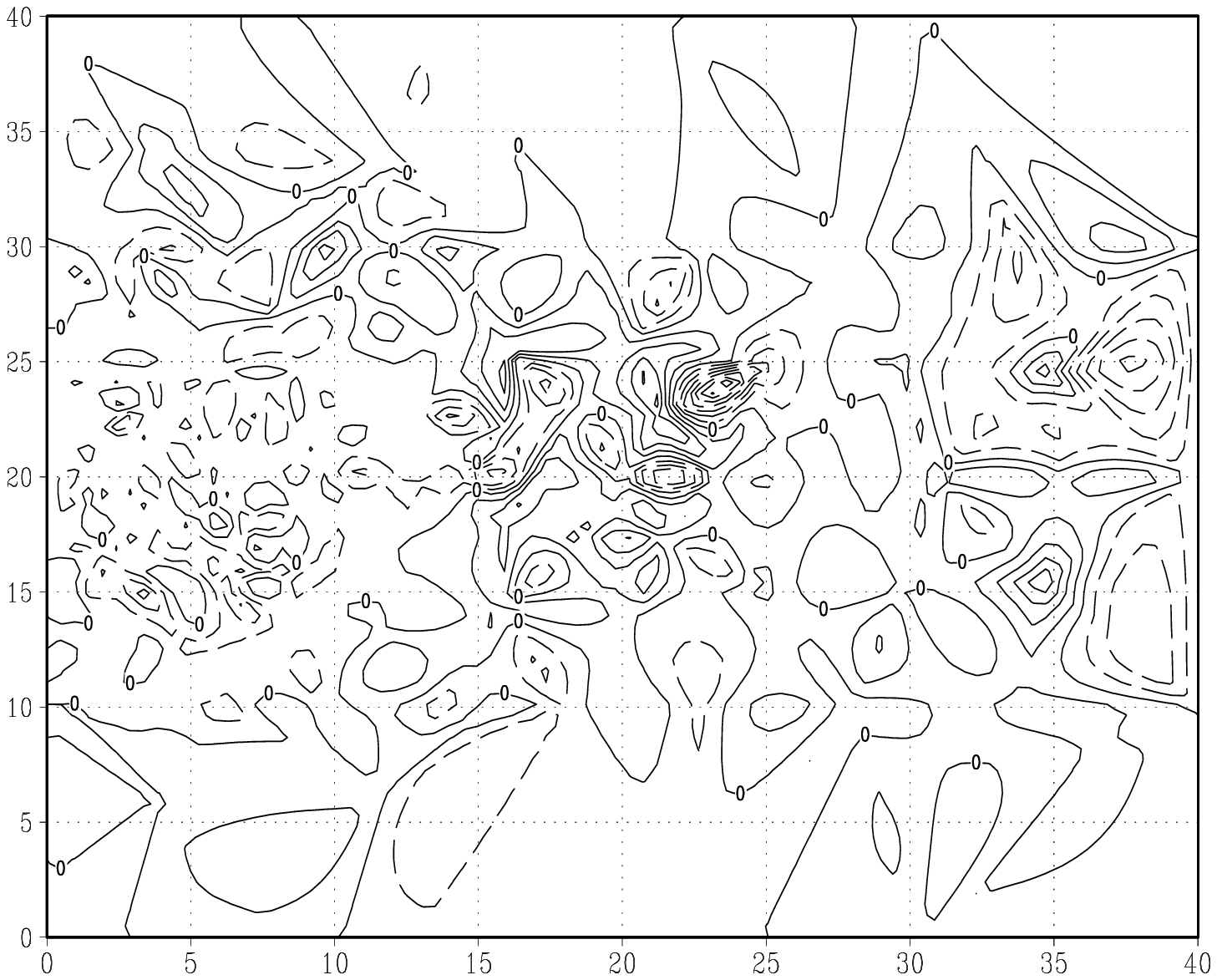}{The most insensitive singular mode for the error growing time $T=12.8$ days. Singular value $4.9\tm 10^{-14}$}
 
Similar effects can be seen in the experiment with the North Atlantic.  
The average most unstable eigenmodes for $T=0.8$ days and for $T=12.8$ days are
shown in \rfg{efunatl}. One can see, that for short error growing time $T=0.8$ the mode is also concentrated in a very small region  represented by  several points on the grid. However, for long error growing time of twelve days, the singular mode occupies a more important region also.

\figureleft{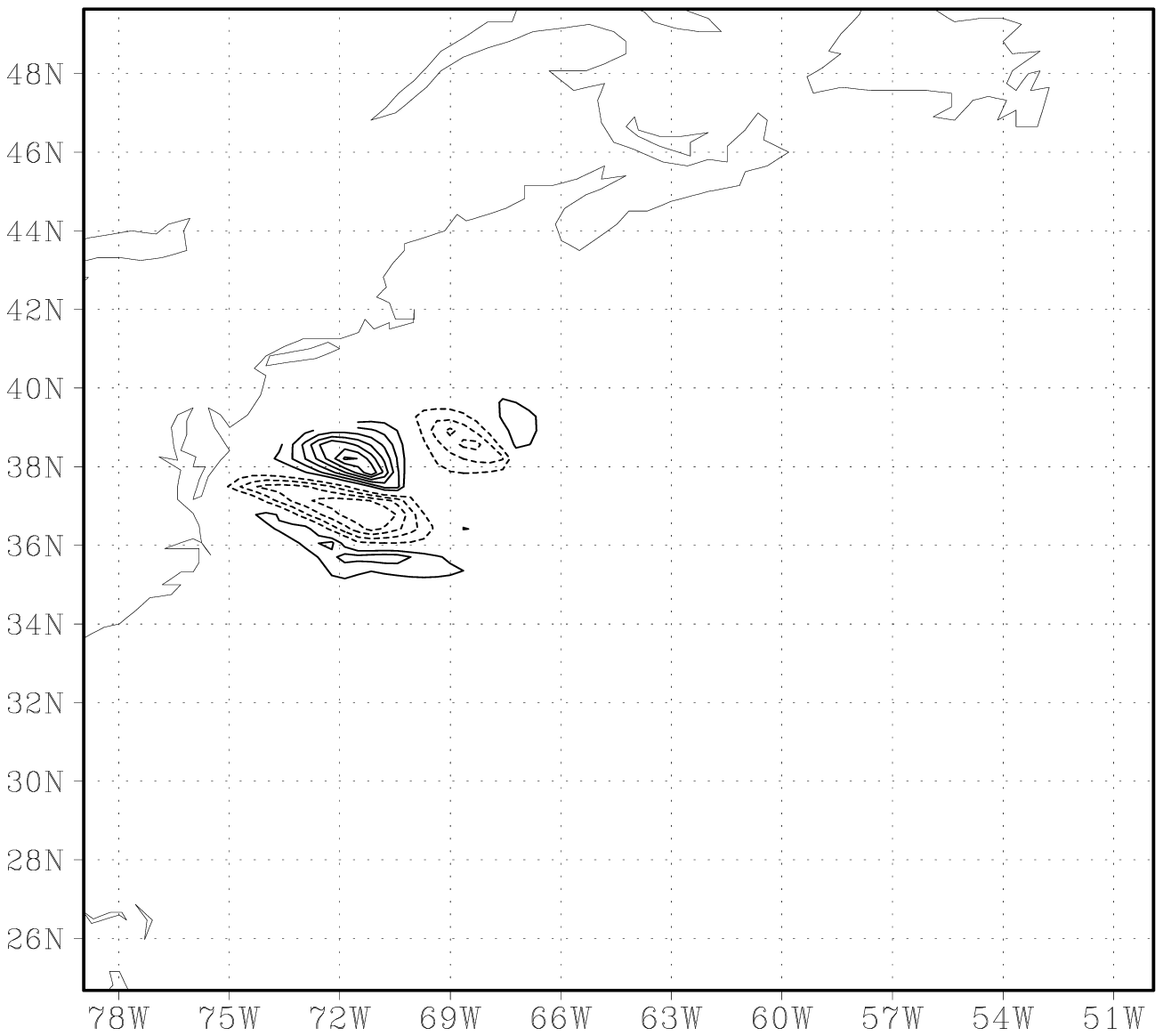}{The most sensitive singular mode in the NA for the error growing time $T=0.8$ days. Singular value $1.6\tm 10^{-8}$}{efunatl}  
\figureright{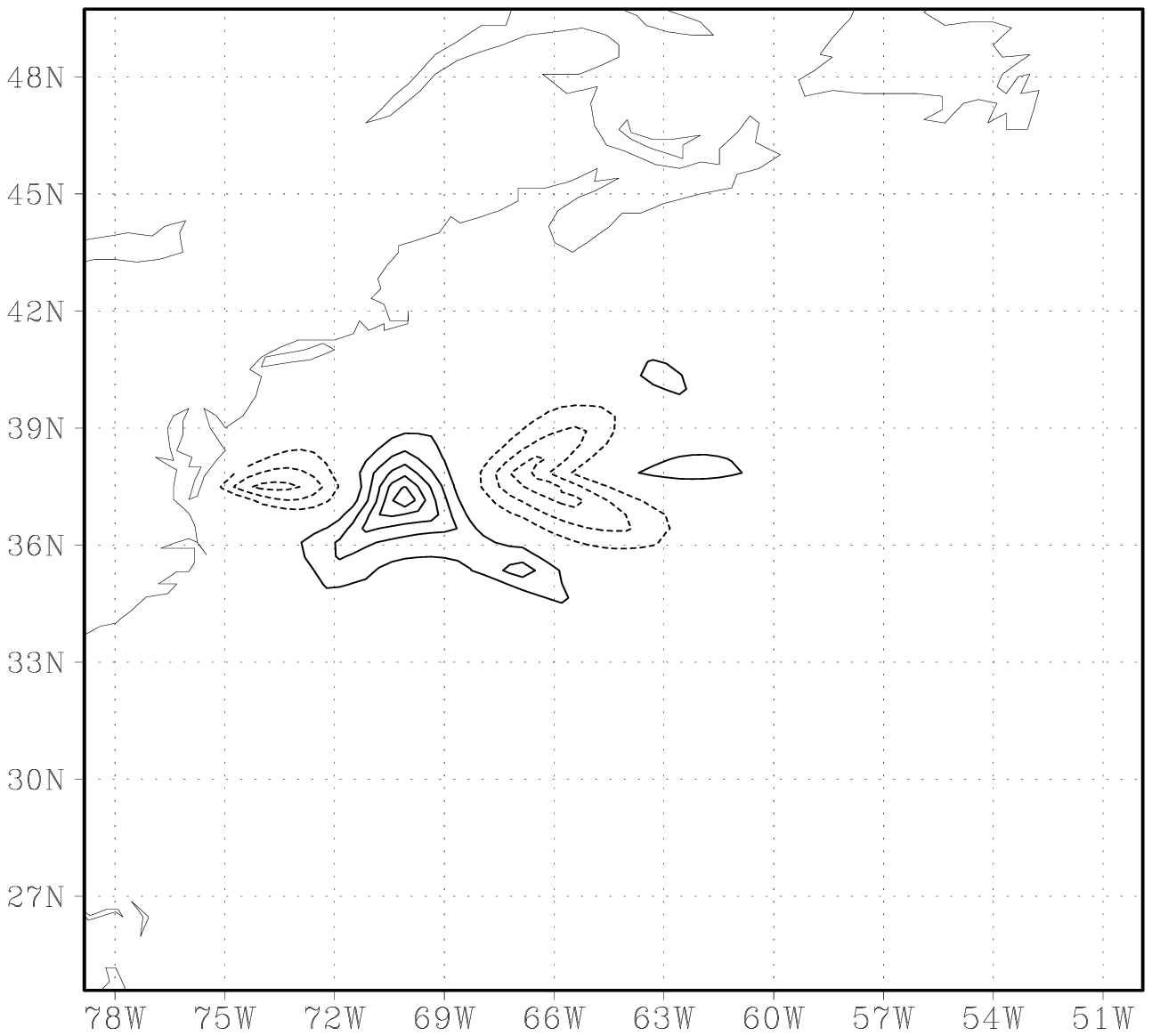}{The most sensitive singular mode in the NA for the error growing time $T=12.8$ days. Singular value $3.3\tm 10^{-5}$}

\figureleft{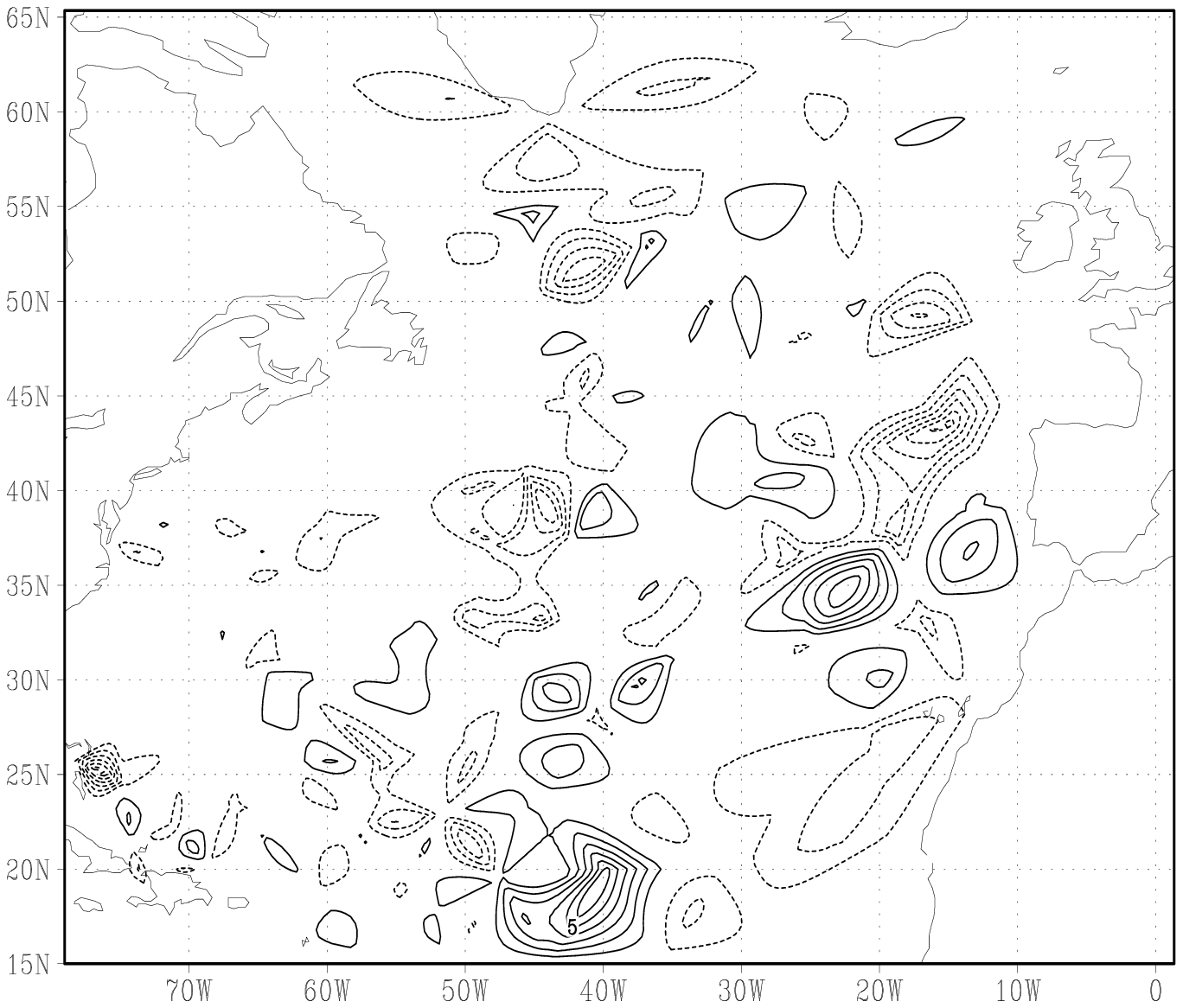}{The most insensitive singular mode in the NA for the error growing time $T=0.8$ days. Singular value $1.6\tm 10^{-15}$}{efunfatl}  
\figureright{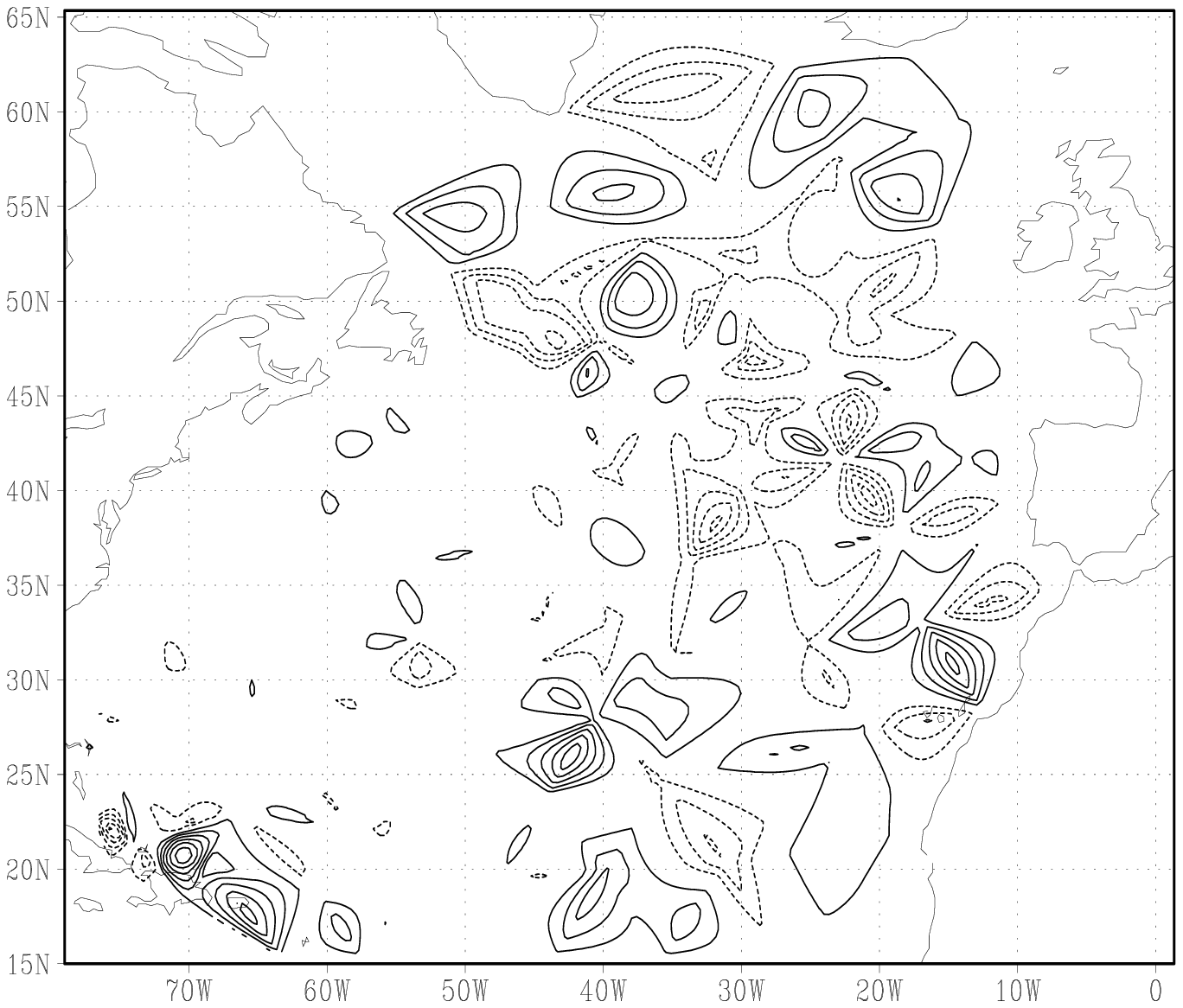}{The most insensitive singular mode in the NA for the error growing time $T=12.8$ days. Singular value $6.5\tm 10^{-15}$}

\subsection{Stationary point}

When the solution is stationary or supposed to be stationary, we can apply the 
formulae \rf{explicit} and avoid  discrete time integration. In this case, 
the computational procedure becomes much easier. Consideration of the stationary point  helps us to see the dependence of singular values on particular parameters, rather than on variations of the basic trajectory.

To get a stationary solution of the model, we increase the dissipation coefficient $\nu$ up to $3000 \fr{m^2}{s}$. The spin-up time required to reach the stationary point is approximately equal to 600 days. We use 800 days spin-up to ensure the stationary behavior of the model after the spin-up. The streamfunction  of this stationary point is shown in \rfg{psistat}. One can see  perfect antisymmetric pattern produced by  antisymmetric wind stress.

\figurecent{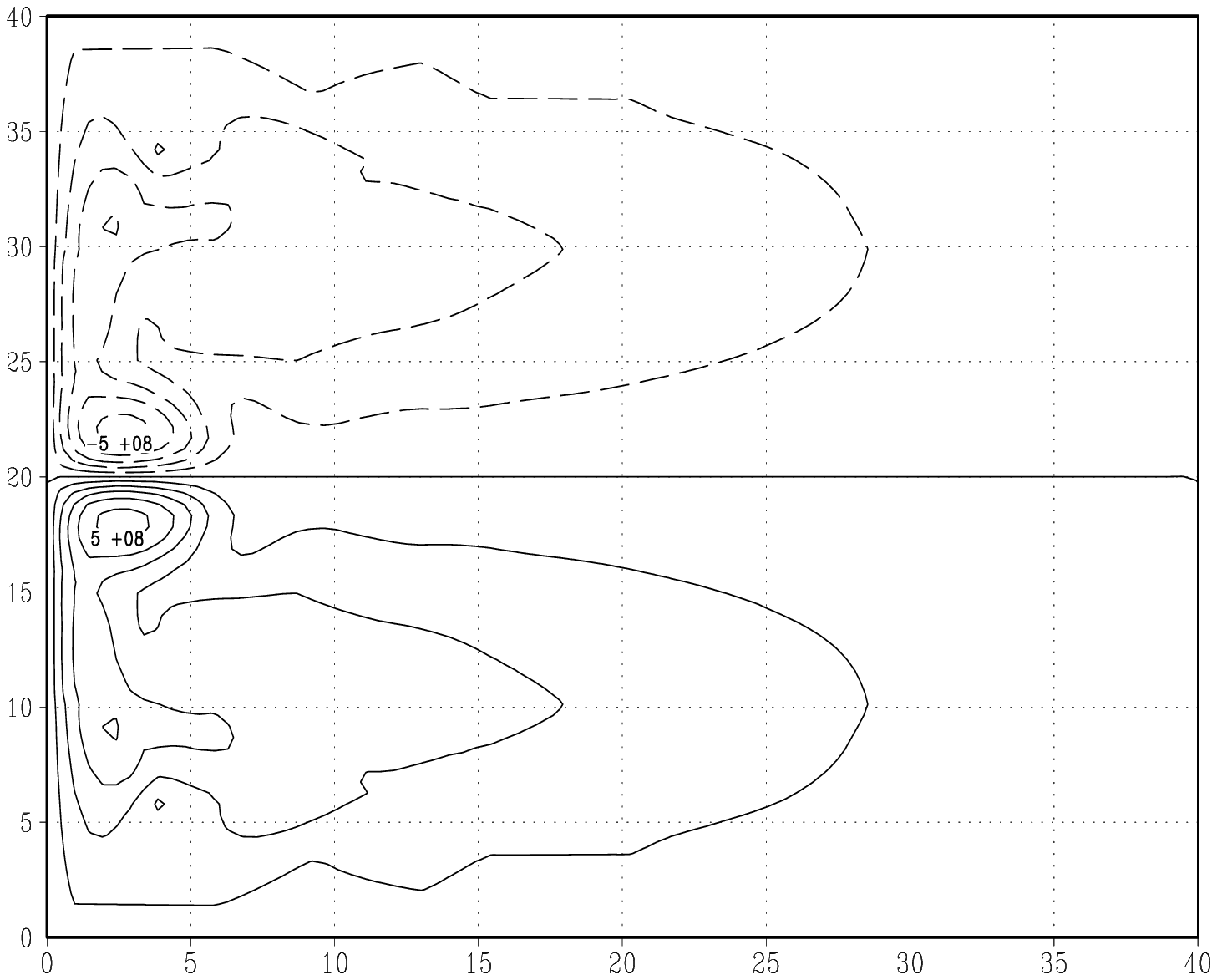}{Streamfunction pattern of the stationary point }{psistat}   

We consider first the dependence of the singular values on the error growing time $T$. As it has been noted above, on large time scales, the dependence must be close to the exponential, but on short time scales, when the source of perturbation is important, the dependence may differ  from an exponential. 
To determine the form of dependence and to identify the tiem scale separating "long" and "short" times, we calculate the eigenvalues of $G^*(T)G(T)$ for the operator $G$  \rf{explicit}:
  \beqnn
 G(T)= (e^{TA}-I) A^{-1} B 
  \eeqnn where operators $A$ and $B$ are defined in \rf{A}, \rf{B}. 
  
\vspace{-0.5cm}
\figureleft{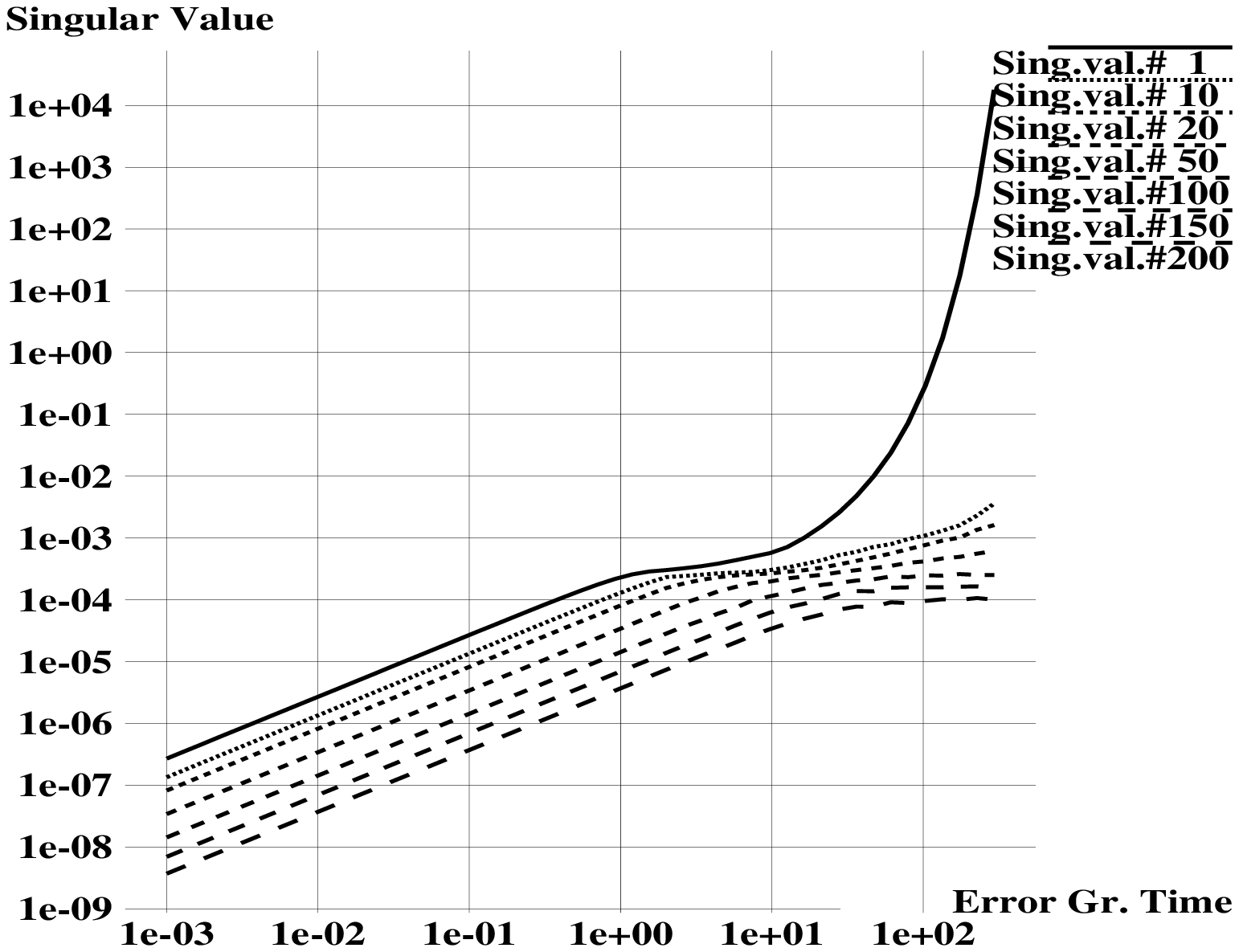}{5  singular values of the matrix $G(T)$ in the square box  versus error growing time $T$ for $10^{-3}<T<100$ days.}{evtevol}  
\figureright{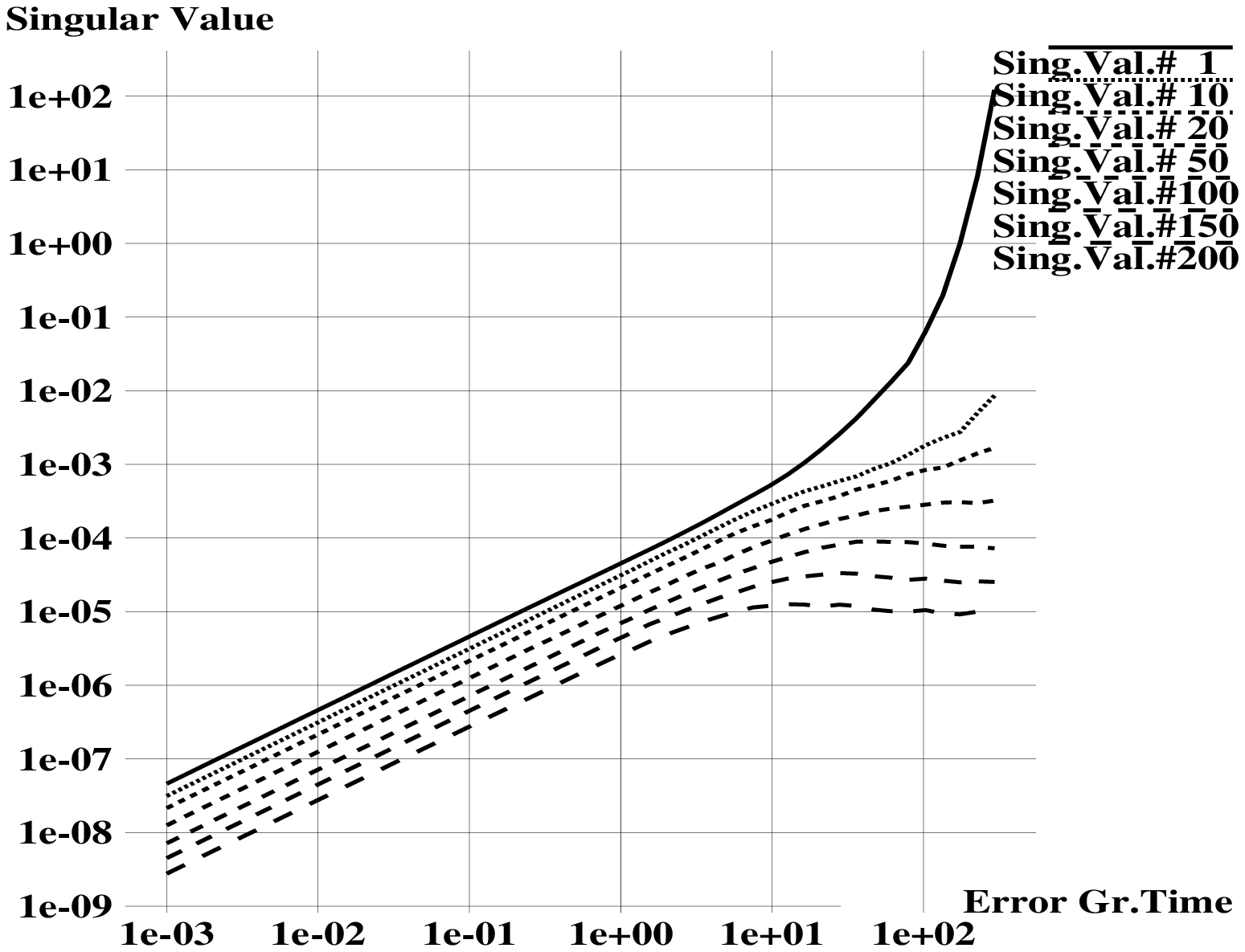}{5  singular values of the matrix $G(T)$ in the North Atlantic versus error growing time $T$ for $10^{-3}<T<100$ days.}

In \rfg{evtevol} we can see the behavior of  5 different singular values of $G$ as the error growing time $T$ increases. The first (largest) value is plotted together with the 10th, the 20th, the 50th, the 100th, the 150th and the 200th.  One can see that  the critical error growing time for all values is close to three days in both experiments. For lower error growing times the increasing rate is close to polynomial because the  rate is linear in logarithmic coordinates
\beqnn 
\ln \lambda = a\ln T +b \mbox{ i.e. } \lambda \sim b T^a 
\eeqnn 
Value of $a$ is equal to 1 in both  experiments. Hence, the  value of $\lambda$ depends  linearly on $T$ when $T$ is lower than the critical value.  The growth of different parts of the spectrum of $G$ is uniform. The lines of the growth of different singular values are parallel to each other.

But, when  error growing time exceeds three days, the growth rate of singular values accelerates and approaches to the exponential.  The growth rate of these  exponentials is very different in different parts of spectrum. In fact, the first singular value increases very rapidly, while the growth of all others is relatively slow. This leads to the fact that the ratio of the largest to the smallest singular value increases   as well as  the small number of the most sensitive singular modes becomes dominant in the whole sensitivity of the model as we have already seen in \rfg{evtsp}. 

Thus, only a few singular values and vectors are important in the sensitivity analysis on long time scales. 

We have seen that   the sensitivity is linear in time on time scales lower than 3 days.  But, on  longer time scales, error growth  becomes exponential. That means, during 2-3 days of the model's time, the perturbation is introduced into the model and, after that, it follows the model's dynamics. During the phase of introduction,  linear transition of perturbation from topography to model's variable dominates, resulting in  linear  dependence on time. But after 2-3 days, it is the model's  dynamics that governs the error evolution. Being non-linear and intrinsicly unstable, the dynamics ensures exponential error growth  of a perturbation. On these time scales, topography perturbation evolves like any other perturbation from any other source.

One of interesting questions is to find how sensitive would be the model with the same parameters but different topography. In particular, it is interesting to see how the sensitivity changes when the topography is not flat but represented by mountains of different height or different space scales. 

 We perform several  experiments with the  same model parameters as above, but with the bottom topography taken as
 \beq
 H(x,y) = 500 \mbox{ m } + \alpha\sin(k_x\pi x)\sin(k_y\pi y) \label{diftopo}
 \eeq 
 The basic state $\psi, \omega$ used to construct the operator $G$ is obtained as the stationary point  reached by the model after the  800 days spin-up. 
 
In the first experiment, the mountains height  $\alpha$ is allowed to vary. We use 30 equally spaced values from $ -300$ to  $ 300$ m: $\alpha_n=-300+ 20\tm n$. Values of $k_x$ and $k_y$ were taken both to be 4.  The topography is shown in \rfg{evcac}B. The perfect sinusoidal form of the topography is somewhat disturbed near  the top and the bottom  sides of the square due to very low resolution at those places (see \rfg{fig1}A).

In this experiment we get  the dependence  of  five  singular values of $G$ on  the bottom mountains height $\alpha$ (\rfg{evcac}A). Error growing time $T$ in this experiment has been chosen as $T=1$ day.  

 \figureleft{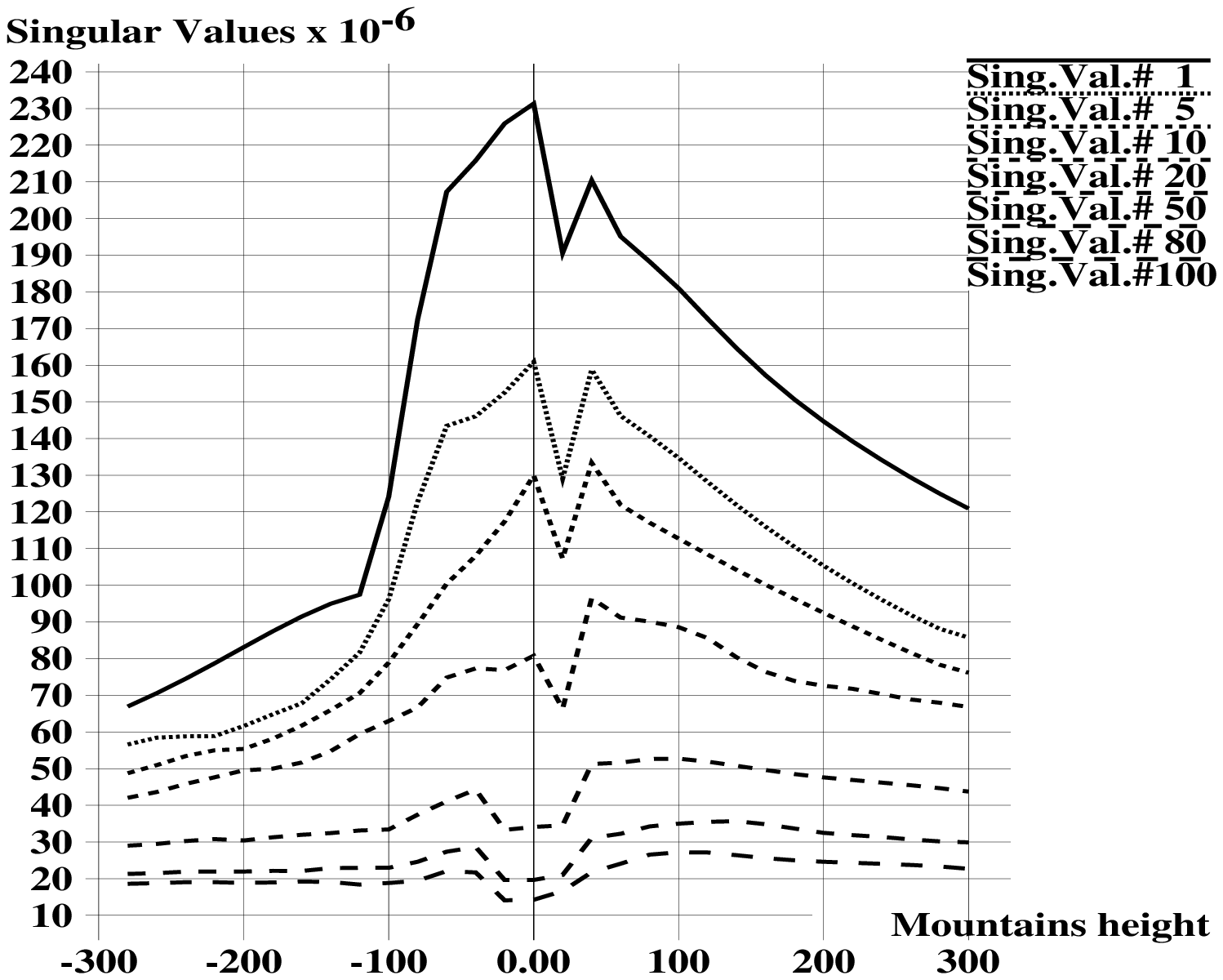}{5  singular values of the matrix $G(1$ day) as functions of mountains height. }{evcac}  
\figureright{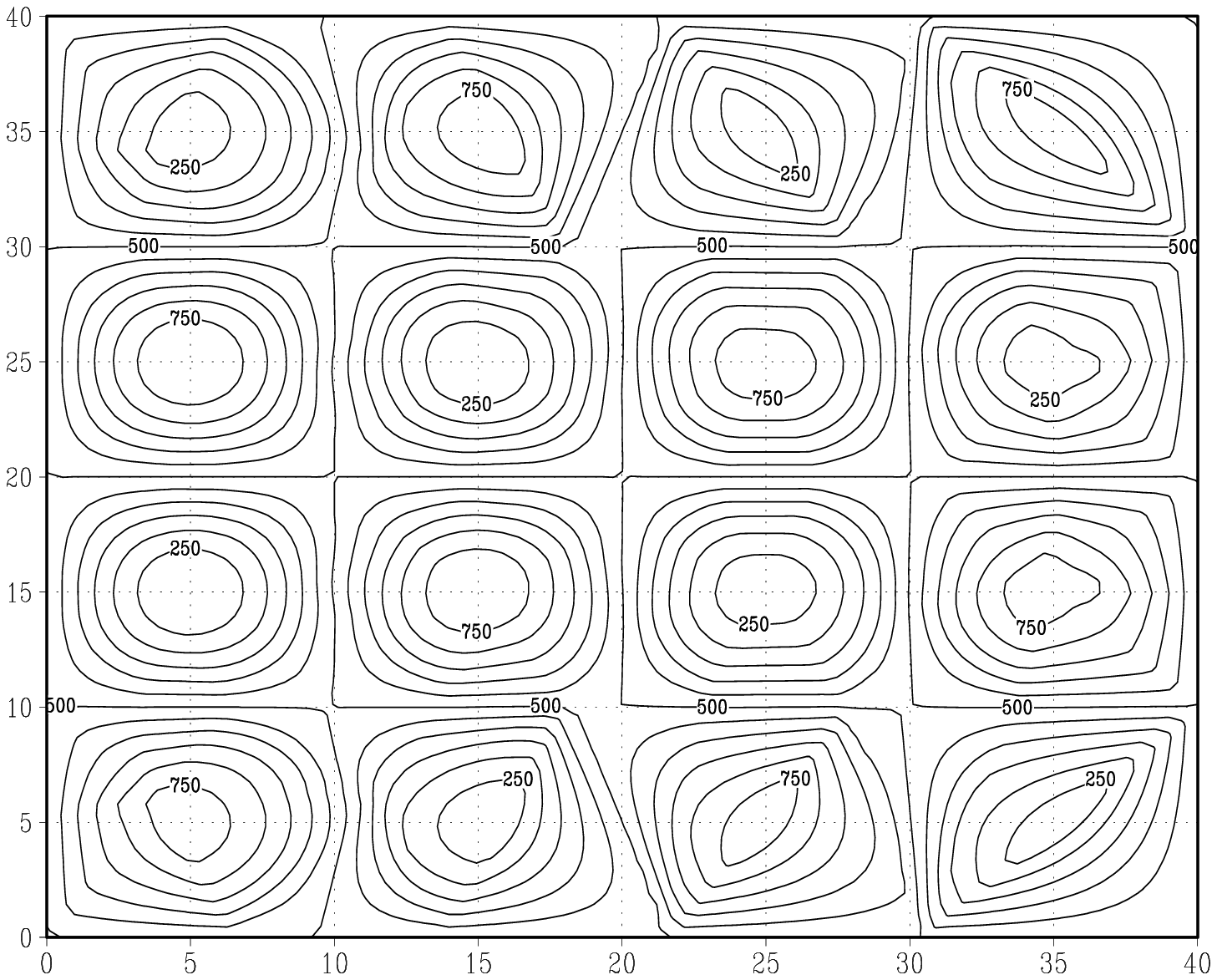}{ Topography in the experiment with variable $\alpha$ }

One can see  the most sensitive circulation is observed when the bottom is flat. If there are some mountains on the bottom, then the sensitivity is lower. However, the  curves in \rfg{evcac}A are not symmetric with respect to zero. When $\alpha$ is negative the largest singular value  is smaller than with the same but positive  $\alpha$.

This difference in behavior can be explained by the streamfunction pattern of the stationary solution obtained for particular topography \rfg{psialpha}. When  $\alpha$ is positive,  topography configuration near the left boundary  anti-correlates  with the streamfunction. That means the vorticity pattern is ``in phase" with the topography. The positive gyre of the streamfunction is situated over the gap on the bottom and amplificated, the negative gyre is over the bottom mountain, and is amplificated also. 

When  $\alpha$ is negative,  streamfunction pattern "correlates" with the topography. The positive anomaly in streamfunction is situated over the hill on the bottom, and the negative anomaly over the gap. This results in lower strength of gyres. Only a little anomaly remains at the usual position in the center of the box, the major part is displaced to the North and to the South, ans is situated at  more favorable topographic position.

\figureleft{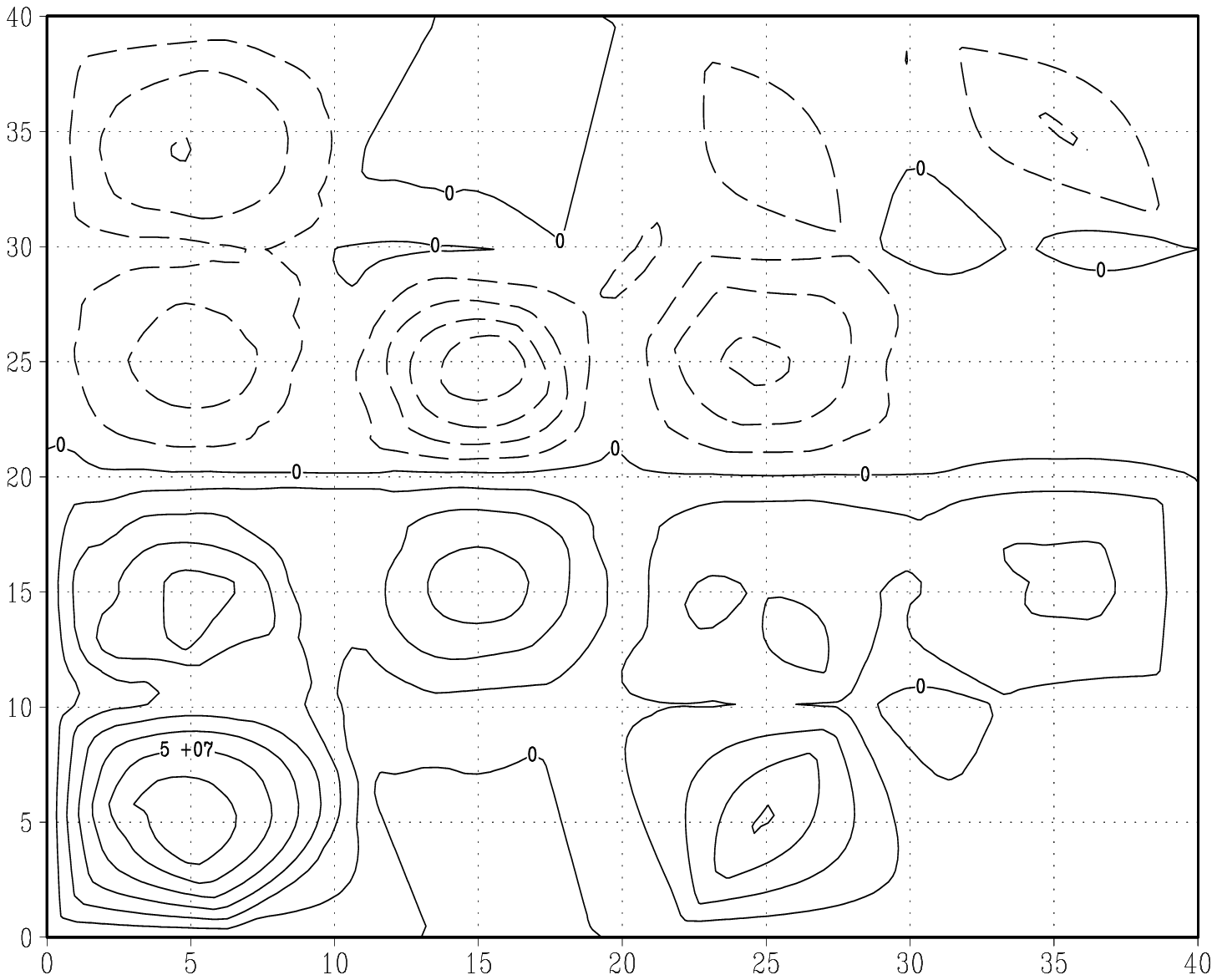}{Basic state streamfunction for  $\alpha=-300$m}{psialpha}
\figureright{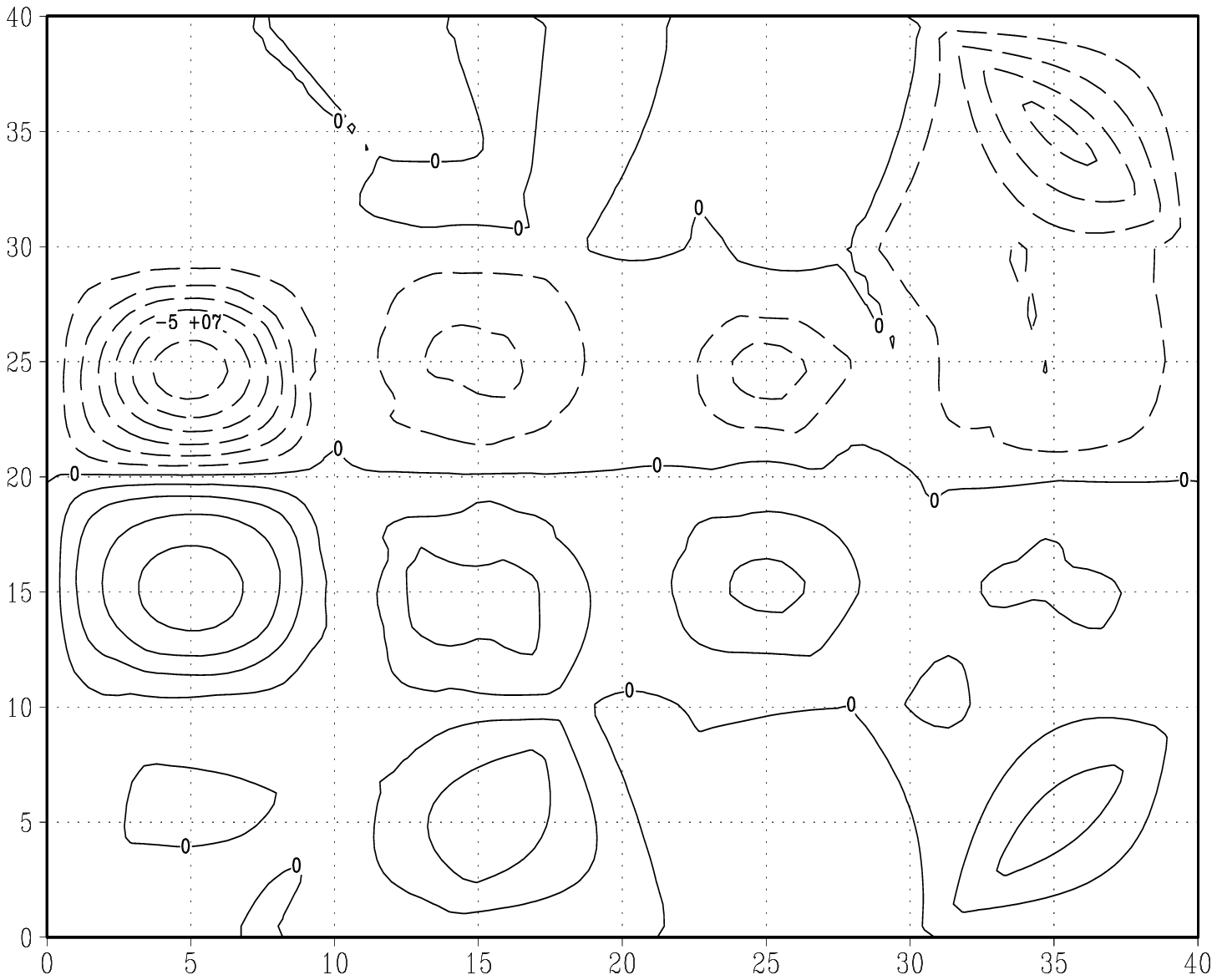}{Basic state streamfunction for  $\alpha=+300$m}

To explain the difference in the sensitivity of the model we may suppose that the sensitivity to the topography is related to the general stability of the model's solution. So far, we  consider  a stationary point of the model, we can evaluate the general stability of the point by the smallest dissipation coefficient $\nu$ in \rf{sw} necessary for the stationary point to exist. It is evident when the dissipation is strong, the model has a stationary point as a global attractor. Any solution starting from any point will tend to the attractor becoming stationary. When the parameter $\nu$ becomes lower, at some value the solution becomes non-stationary. It is this value that we compare with the largest singular value of the matrix $G$ in \rfg{ev-mu}. One can see a clear relationship between the stability and the sensitivity of the solution.

\figurecent{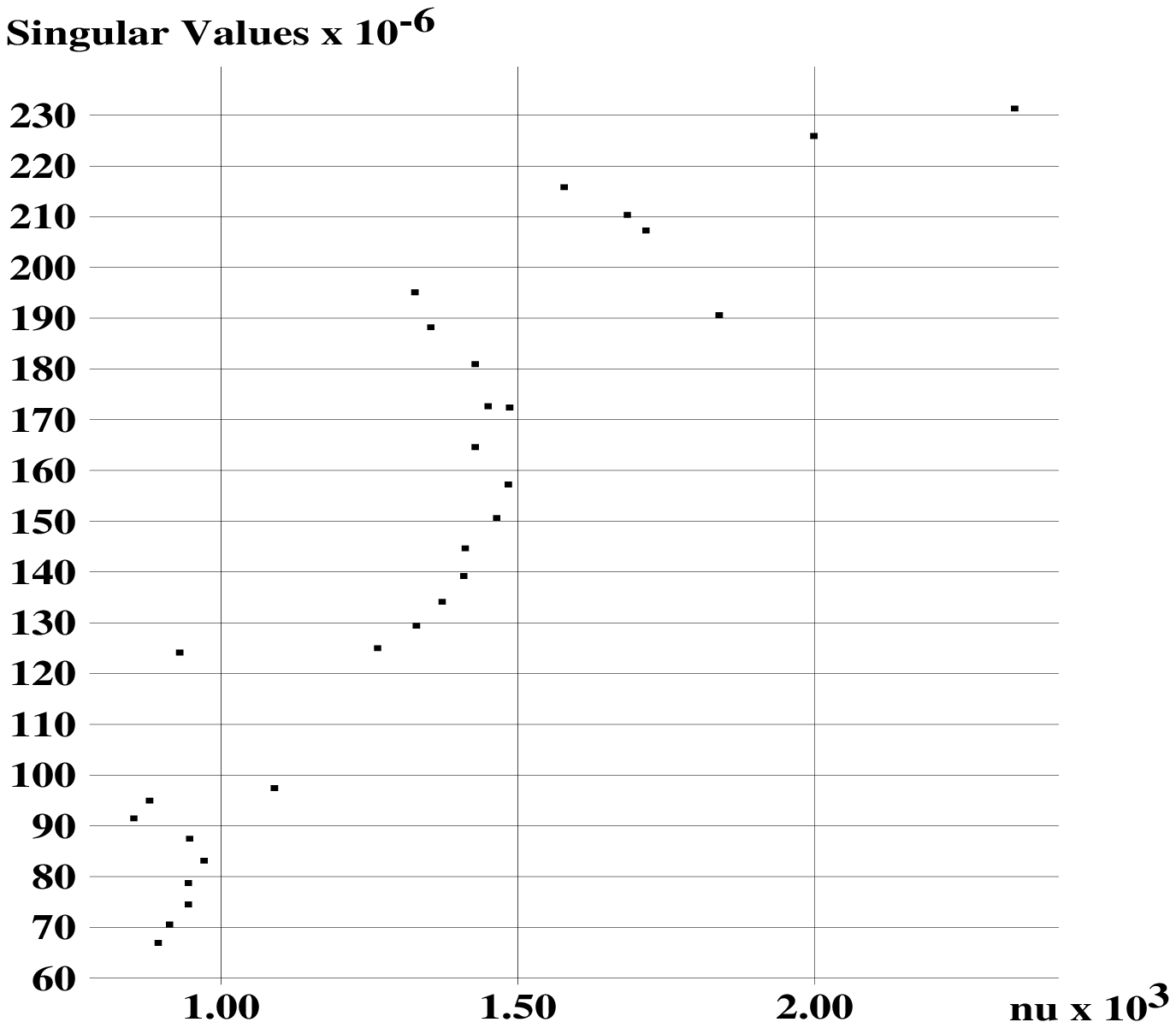}{The smallest $\nu$ of the stationary point vs largest singular value of the matrix $G$}{ev-mu}

In the second experiment we look at the sensitivity of the model with the sinusoidal  topography of different wavenumbers. We use the same formula \rf{diftopo} for $H(x,y)$ with  constant $\alpha=100$m but with varying wavenumbers $k_x$ and $k_y$  from 0 (flat bottom)  up to 30 (short scale mountains, at the limit of  grid resolution).  Topography pattern for $k_x=k_y=10$ and $k_x=k_y=30$ are shown in \rfg{topof}. One can see the topography with the wavenumber 10 is well resolved in the whole center of the domain, while the wavenumber 30 is only possible to resolve in the small region near the jet-stream.

\figureleft{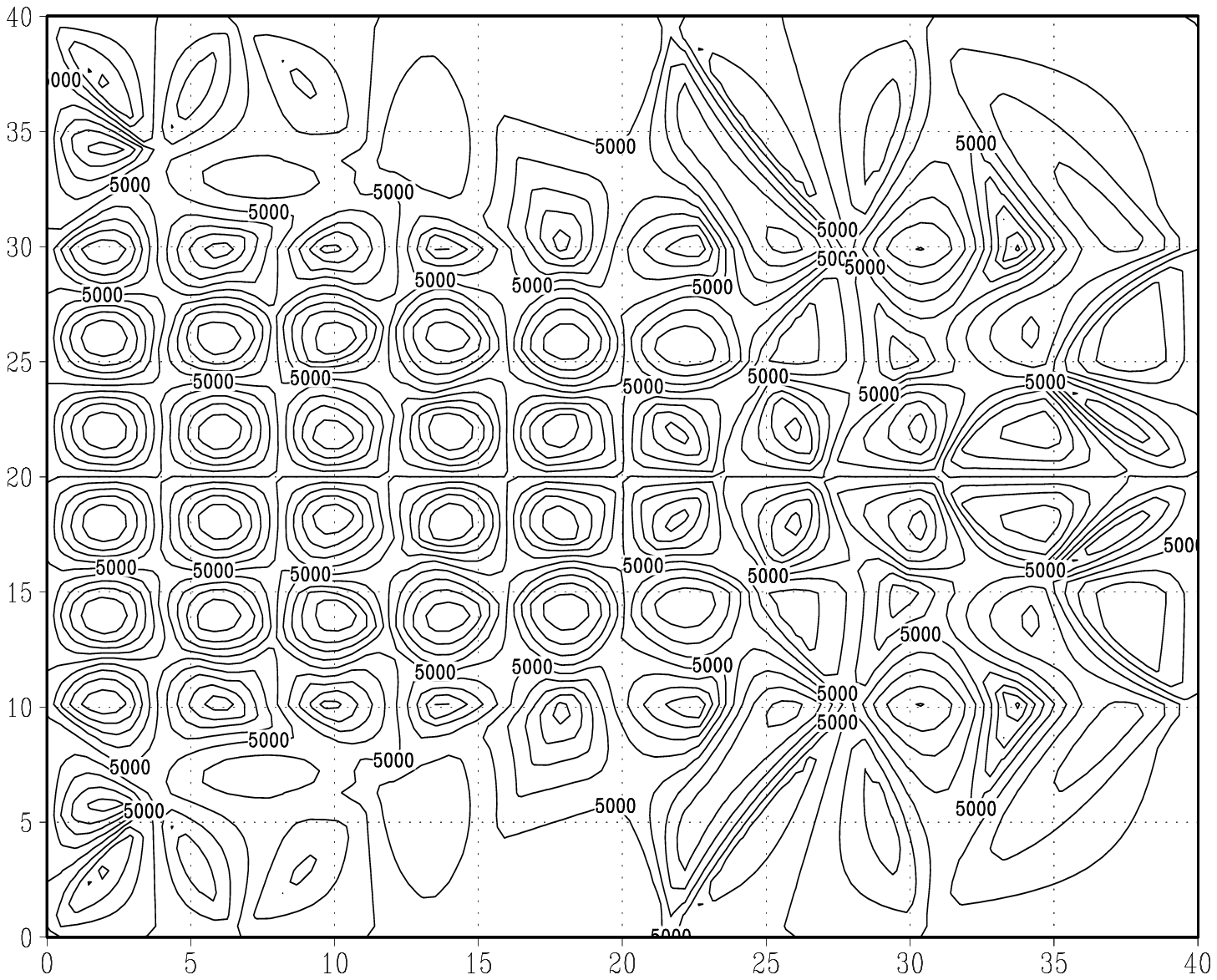}{Topography pattern with $k_x=k_y=10$}{topof}
\figureright{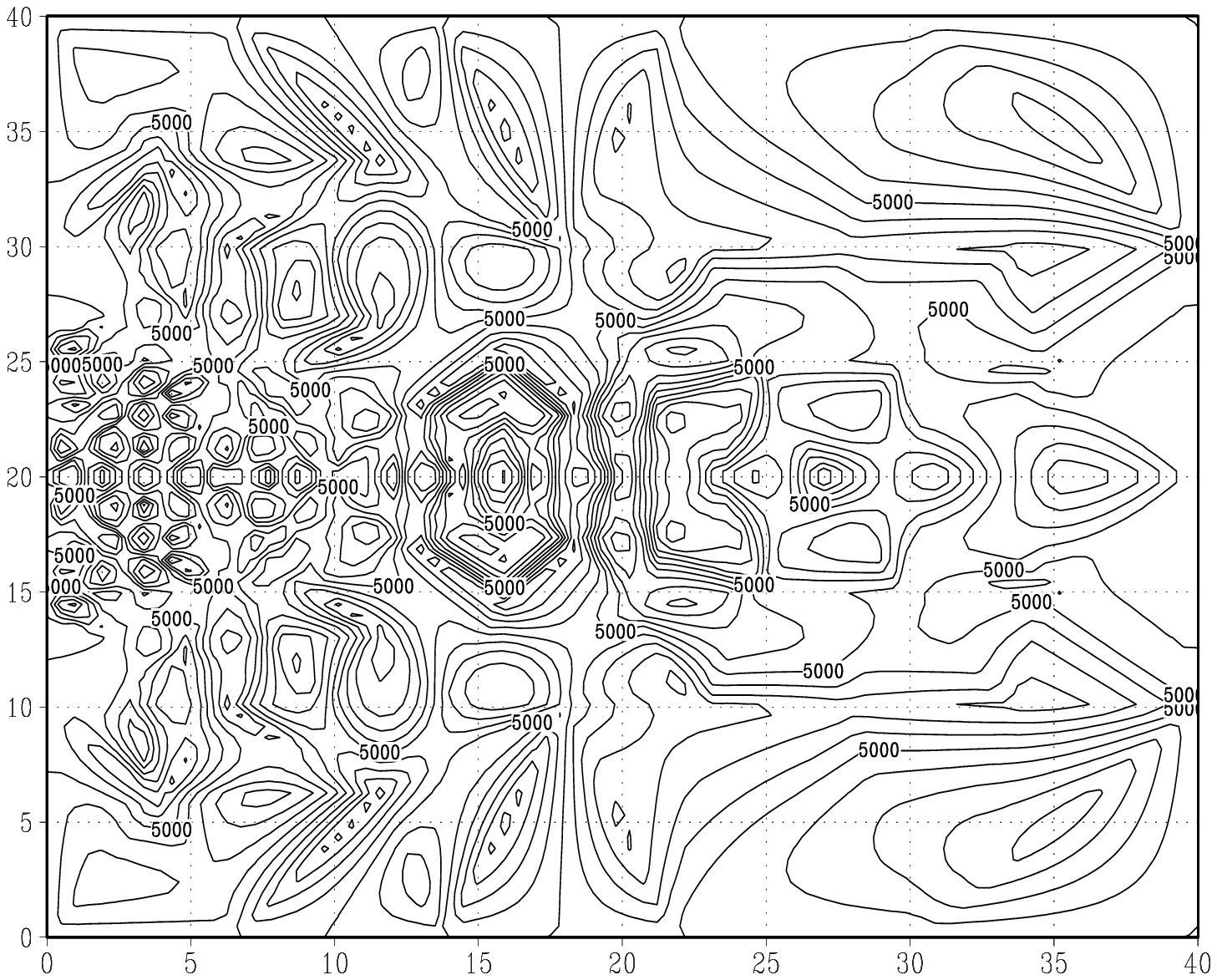}{Topography pattern with $k_x=k_y=30$}

Streamfunction patterns   of stationary points for these two topographies are presented in \rfg{psif}. We can see both patterns are no longer antisymmetric as seen in \rfg{psistat}. However, streamfunction's deformation in this experiment  is not as drastic as in the experiment with the amplitude variations \rfg{psialpha}. The jet-stream is concentrated at the usual place and its intensity remains almost the same.  

\figureleft{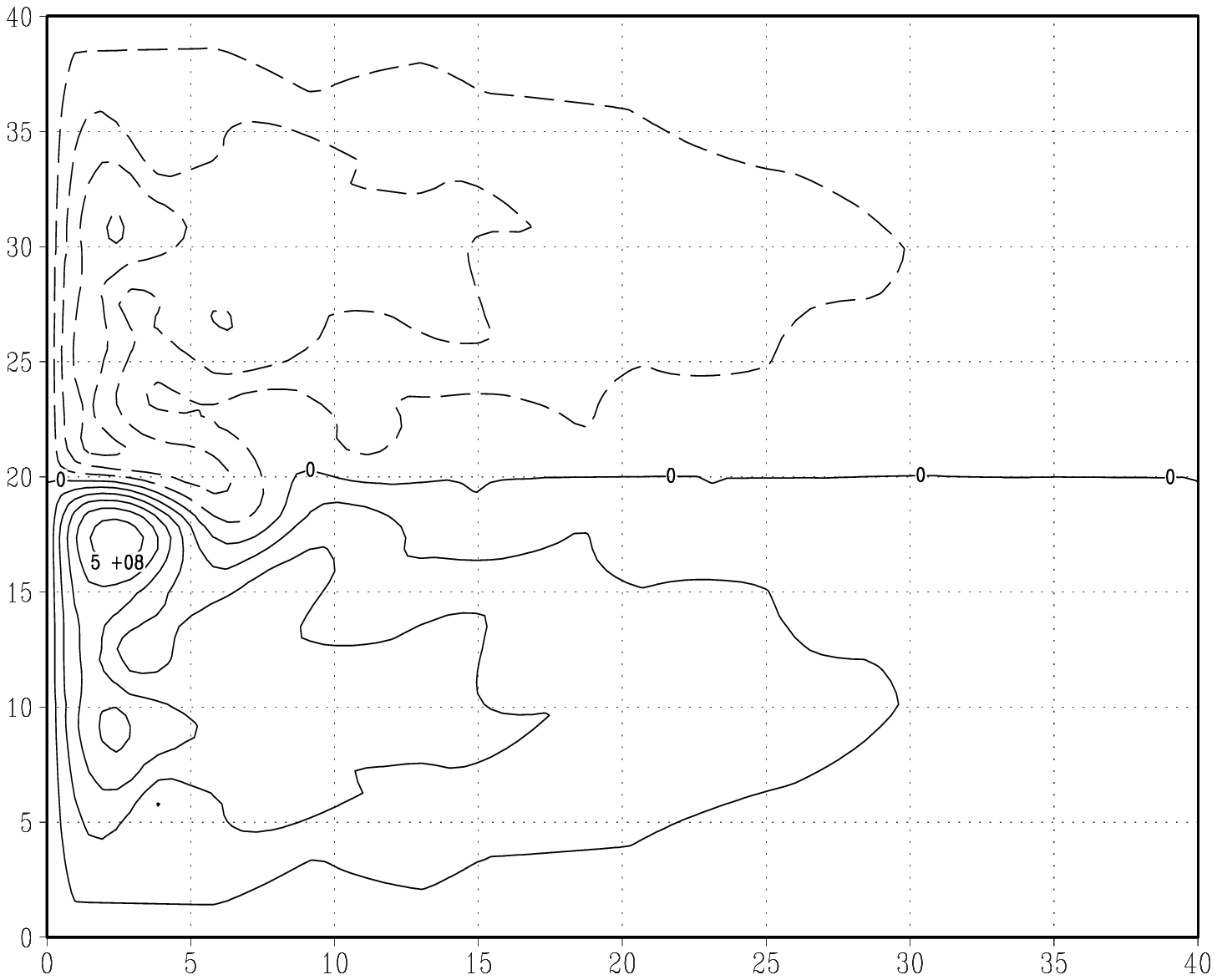}{Basic state streamfunction for  $k_x=k_y=10$}{psif}
\figureright{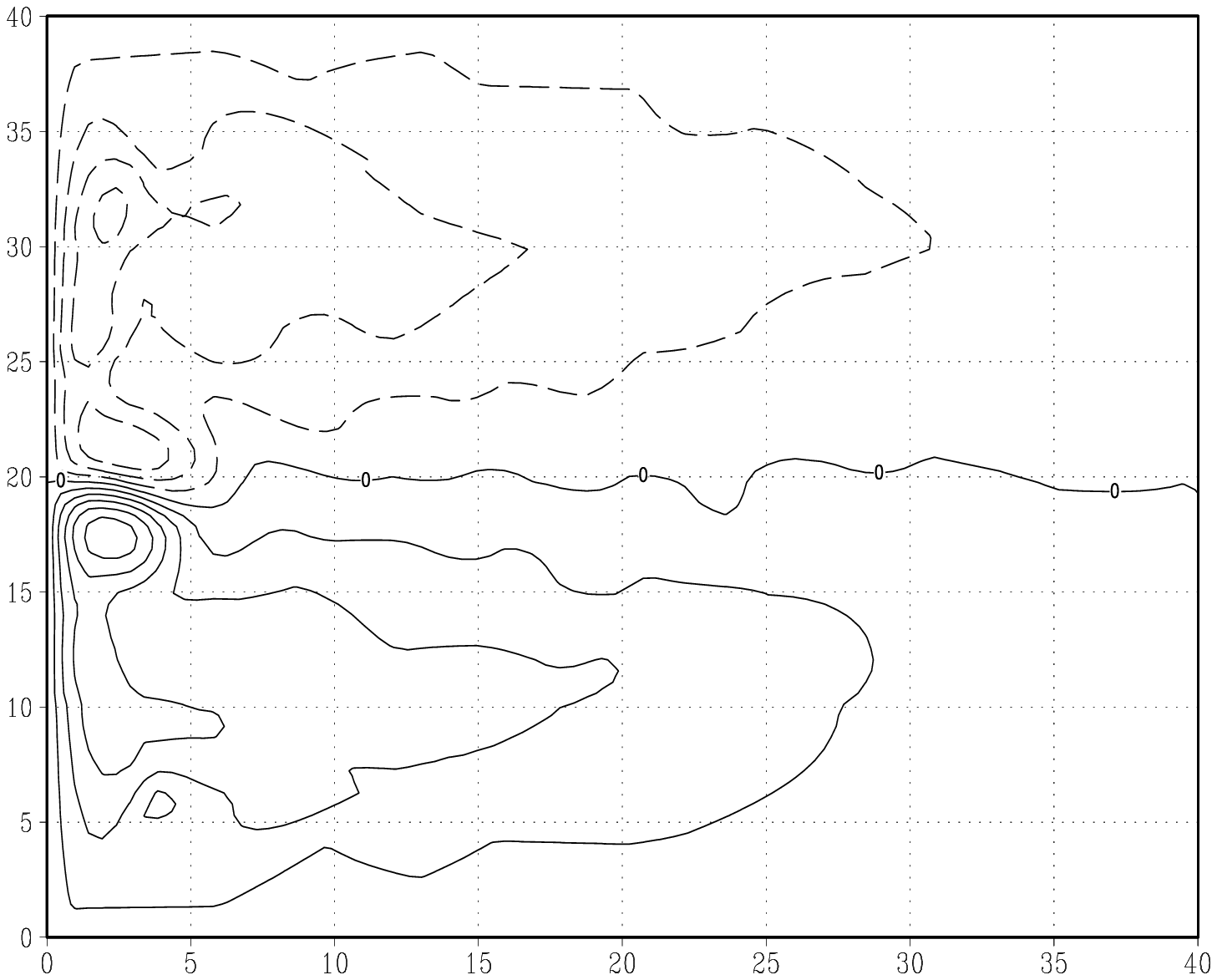}{Basic state streamfunction for $k_x=k_y=30$}

The dependence of singular values on the topography's wavenumber is presented in \rfg{evevolf}. One can see that variations of singular values are  smaller  than in the experiment with the amplitude change. In addition, there is no tendency when  wavenumbers increase. The behavior of singular values exhibits irregular variations about  values of sensitivity of the model  with  flat bottom.  The amplitude of variation is relatively small, remaining in frames that the maximal value is less than two times the minimal one. In fact, these variations are similar to natural variations of the sensitivity due to temporal evolution of the circulation with flat bottom. We can not distinguish a particular tendency due to the rugosity of the topography.

\figurecent{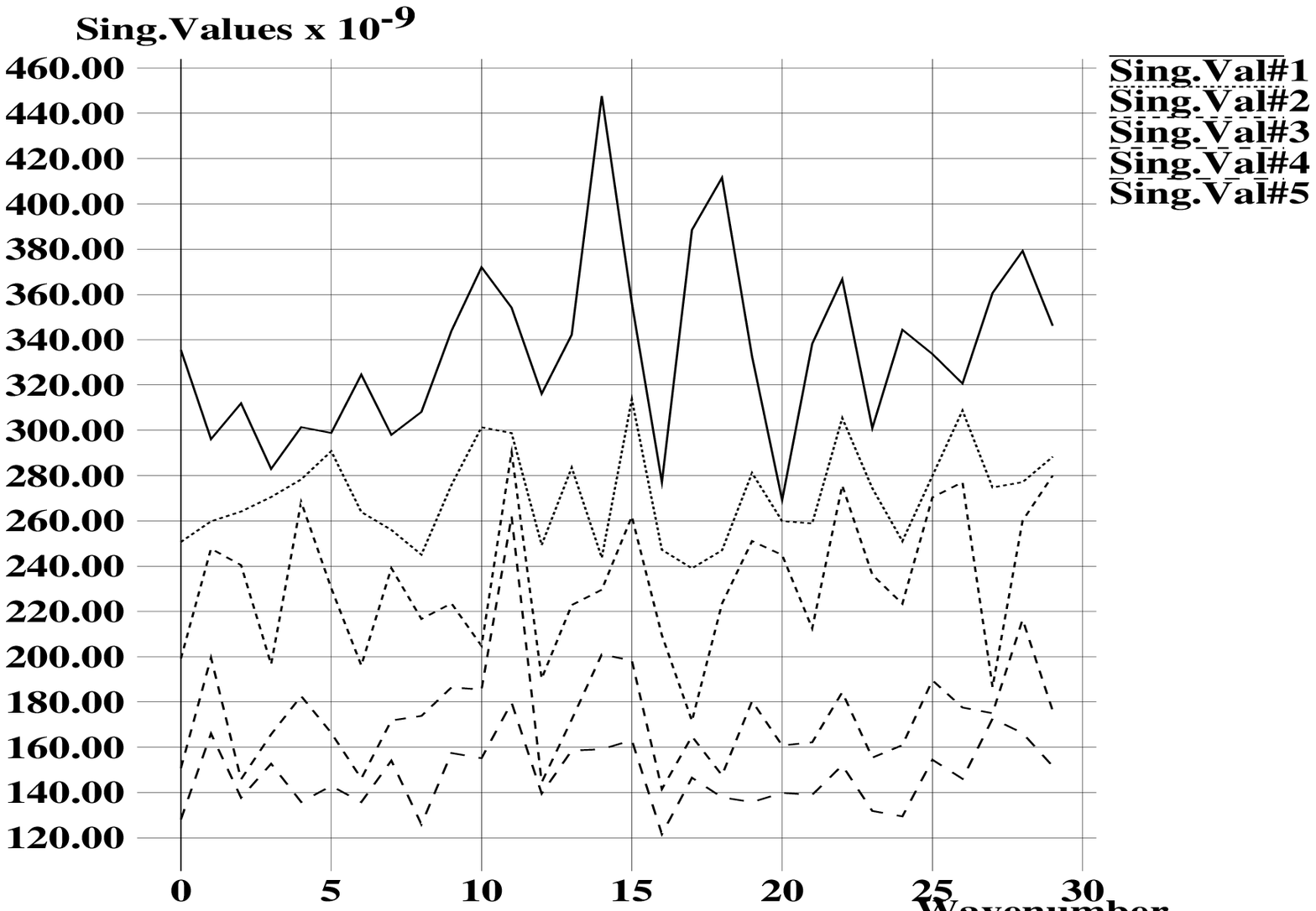}{ Evolution of the 5  singular values of the matrix $G(T)$ for the model with the topography with different wavenumbers }{evevolf}

\section{Conclusion}
We have considered the sensitivity of the vorticity field to the topography perturbations in  frames of the barotropic ocean model.  Both stationary and non-stationary solutions of  nonlinear model have been viewed. 

We distinguish the analysis of the quantitative measure of the sensitivity, expressed in singular values of the operator, and the qualitative pattern of the singular function, that precise the geographical region of particularly sensitive or insensitive solution. 

The quantitative measure is  influenced  essentially by  error growing time. Longer is the time period during which we allow the error to grow, greater the sensitivity is. This conclusions is consistent with numerous studies of the model's sensitivity to other parameters. Thus, the predictability studies that analyze the sensitivity to initial conditions, reveal the exponential (or close to) growth rate. Regarding to this, one can cite  \cite{Barkmeijer}, \cite{LT}, \cite{MP93}, \cite{Nicolis95}, \cite{kaz99} and many others. In this paper we show, that at short time scales, the sensitivity to topography  differs from the sensitivity to initial conditions. But, in the long time limit, the sensitivity of the solution is the same to any source of perturbation. The intrinsic model's instability dominates at these time scales and the source of the perturbation is no longer important. 

We have analyzed  patterns of the most sensitive modes of the solution. These patterns point out the regions where the solution more sensitive to the topography perturbations. The sensitivity is important in regions where the flow is turbulent. On the other hand, in regions where  the flow is laminar, the solution exhibits lower sensitivity to the topography.  The conclusion is in agreement with the result of \cite{LoschWunsch}, where it was shown that the sensitivity is largest where current speeds are high. 

Barotropic model in the North Atlantic develops  turbulent flow near  American coast and laminar flow near the European one. The  sensitivity of the model's solution to topography is low in this region. All sensitivity modes corresponding to small eigenvalues are concentrated in this region. That means,  the data assimilation procedure intended to  reconstruct  the topography in this region may not be efficient because the influence of corresponding singular modes on  the model's solution is small. One should use additional apriori  information to reconstruct topography.

Turning attention to more realistic models, a number of problems can be encountered. First of all this concerns a multi-layer model with different geometry in each layer. The presence of baroclinic component may also change the sensitivity of the model. Second, this study gives no information about particular schemes of parametrisation of topography. Numerous modern schemes like partial step  or  shaved cells can not be distinguished in this paper as well as different grids like Arakawa's ones. Working with the barotropic vorticity equation  we can not pay an attention to, for example, C-grid, which is frequently used in realistic models.

Acknowledgments. All the contour pictures have been prepared by the Grid Analysis and Display System (GrADS) developed in  the Center for Ocean-Land-Atmosphere Interactions,   Department of Meteorology, University of Maryland.

\section{Appendix: Numerical resolution}

        In order to look for a weak solution of the problem ~\rf{mateq} we perform  its variational formulation:
\beq
\spm{\der{\delta\omega}{t},\varphi(x,y)} =   \spm{A(\psi,\omega)\delta\omega,\varphi(x,y)}+ 
  \spm{B(\psi,\omega)\fr{\delta H}{H},\varphi(x,y)} \label{varform}
\eeq
for any function  $\varphi(x,y) \in H^1_0(\Omega)$. Here, $H^1_0(\Omega)$ denotes  the linear space of functions 
that the square is integrable as well as the square of their first derivatives. Functions in this space must  vanish on the boundary of the domain. Brackets $\spm{.,.}$ denote the $L_2$ scalar product:
\beq
\spm{\psi,\varphi}=\intint_\Omega \psi\varphi dxdy \label{sp}
\eeq

The variational formulation ~\rf{varform}
 of the problem ~\rf{mateq} allows us to search the solution by the  finite element method (FEM). 

So far the  model \rf{btp} under consideration is similar to the barotropic one and the solution produced by the barotropic model of the North Atlantic typically includes a western boundary layer with intense velocity gradients, the advantage of refining the triangulation along the western boundary of the domain is rather clear. This helps to keep the quality of explicit eddy resolution by the model while working with lower number of grid nodes. The comparison of finite elements (FE) and finite difference (FD) models performed in \cite{LPBB} revealed that the difference arose between simulations by FE and FD techniques can be judged as insignificant when the number of FE nodes is about 6 times lower than the number of FD ones. 

In spite of the fact that the number of operations per time step and grid node is much higher for FE model, the possibility of reducing the number of grid points considerably diminish the computational cost of a model run.  The possibility to have a better precision working with a lower number of grid points is very valuable in this work  allowing us to perform more detailed study of the sensitivity.

The package MODULEF \cite{MODF} has been used to perform a triangulation of a domain. This package produces quasi-regular triangulation of the domain   basing on the prescribed grid nodes on its boundary.  We require the refining of the triangulation near the western boundary and especially in the middle of the domain where velocity gradients are extremely sharps.

 The domain $\Omega$ is covered by  a set of non-intersecting triangles. The set of integration points is defined as the union of vertices and mi-edges of triangles. 
Finite elements of type $P_2$ are used here, i.e. the polynomials of the second degree $p_i(x,y)=a_i x^2+b_i xy+c_i y^2+d_i x+e_i y+f_i$. The $i$th finite element is taken to be
equal to 1 at the i-th integration point and zero at all other points.

According to the Dirichlet boundary conditions \rf{bcpert}, we consider internal points of the domain only: $(x_i,y_i) \in \Omega \backslash \partial\Omega \mbox{ for } i = 1,\ldots,N. $, so variables of the problem  are presented as linear combinations
\beqr
 \psi(x,y,t) = \sum\limits_{i=1}^{N} \psi_{i}(t) p_i(x,y),  \nonumber\\
 \fr{\omega(x,y,t)+ f_0+\beta y}{H(x,y)} = 
 \sum\limits_{i=1}^{N} \biggl(\fr{\omega+ f_0+\beta y}{H}\biggr)_i p_i(x,y) \nonumber\\
 \fr{\delta H(x,y)}{H(x,y)} = \sum\limits_{i=1}^{N} \biggl(\fr{\delta H}{H}\biggr)_{i} p_i(x,y),  \hspace{3mm}
 \delta\omega(x,y,t) = \sum\limits_{i=1}^{N} \delta\omega_{i}(t) p_i(x,y) \label{develfun}
\eeqr 
To simplify notations, we define matrices of mass and rigidity as
\beqr
{\cal M}_{i,j} = \hspace{2mm} <p_i,p_j>, \hspace{5mm}
{\cal C}_{i,j} = \hspace{2mm} <\nabla p_i,\nabla p_j> \hspace{5mm}
\left\{
\begin{array}{rl} i&=1,\ldots,N \\
                  j&=1,\ldots,N
\end{array}  \right.
\eeqr

We multiply the equation \rf{mateq} by finite elements $p_k(x,y), \; \forall k=1 \ldots N$.
\beqr
\spm{\der{}{t}\sum\limits_{i=1}^{N} \delta\omega_{i}(t) p_i(x,y),p_k(x,y)} =   \spm{A(\psi,\omega)\delta\omega,p_k(x,y)}+ \nonumber\\
  \spm{B(\psi,\omega)\fr{\delta H}{H},p_k(x,y)} \label{mateqpj}
\eeqr
Scalar products with operators $A$ and $B$ are developed as follows. From \rf{A} we get 
\beqr
\spm{A(\psi,\omega)\delta\omega,p_k} &=& 
-\spm{\jac(\psi,\fr{\delta\omega}{H}),p_k} +
\spm{\jac( \fr{\omega+ f_0+\beta y}{H}, \biggl(\nabla \fr{1}{H}\nabla\biggr)^{-1} \delta\omega ),p_k} -\nonumber \\ &-&
\nu\spm{\nabla\delta\omega,\nabla p_k} -\sigma\spm{\delta\omega, p_k} \label{atmp}
\eeqr
using \rf{develfun} we write the first Jacobian of \rf{atmp} as
\beq 
\spm{\jac(\psi,\fr{\delta\omega}{H}),p_k}  =
  \sum\limits_{i=1}^{N}\sum\limits_{j=1}^{N} \psi_{i}(t)\biggl(\fr{\delta\omega}{H}\biggr)_j \spm{\jac(p_i,p_j),p_k} 
\label{a1}
\eeq
and the second one as 
\beq  
\spm{\jac( \fr{\omega+ f_0+\beta y}{H}, \biggl(\nabla \fr{1}{H}\nabla\biggr)^{-1} \delta\omega ),p_k} = 
 \sum\limits_{i=1}^{N}\sum\limits_{j=1}^{N}
 \biggl(\fr{\omega+ f_0+\beta y}{H}\biggr)_i \xi_j \spm{\jac(p_i,p_j),p_k} 
 \label{a2}
 \eeq
 where $\xi$ is determined as the solution of the equation 
 \beqr
  \biggl(\nabla \fr{1}{H}\nabla\biggr)^{-1} \delta\omega &=&\xi \nonumber \\
  \nabla \fr{1}{H}\nabla\xi &=& \delta\omega. \nonumber
 \eeqr
 To solve this equation, we multiply it  by $p_k(x,y), \; \forall k=1 \ldots N$ and integrate by parts. That  gives
\beqnn
-\spm{\fr{1}{H}\nabla\xi,\nabla p_k }=\spm{\delta\omega ,p_k}
\eeqnn
Supposing $\xi$ to be discretised in the same way as other functions $\xi(x,y,t) = \sum\limits_{i=1}^{N} \xi_{i}(t) p_i(x,y)$, we get 
 \beqnn
 \sum\limits_{i=1}^{N}\sum\limits_{j=1}^{N} \biggl(\fr{1}{H}\biggr)_i \xi_j \spm{p_i\nabla p_j,\nabla p_k}
 =
 \sum\limits_{j=1}^{N}  \spm{p_i,p_j} \delta\omega_j 
 \eeqnn
 or in matricial form 
 \beq
 {\cal H} \xi = {\cal M} \delta\omega \mbox{ where } {\cal H}_{i,j} = \sum\limits_{k=1}^{N} \biggl(\fr{1}{H}\biggr)_k  \spm{p_k\nabla p_j,\nabla p_i}
\label{math}
\eeq
Combining  \rf{atmp}, \rf{a1}, \rf{a2} and \rf{math} we get
\beqr 
\spm{A(\psi,\omega)\delta\omega,p_k} &=&  (A^{(1)}+A^{(2)}{\cal H}^{-1}{\cal M} -\nu{\cal C} -\sigma{\cal M})\delta\omega 
\label{amat} \\ 
A^{(1)}_{k,j}&=&\sum\limits_{i=1}^{N} \psi_{i}(t)\biggl(\fr{1}{H}\biggr)_j \spm{\jac(p_i,p_j),p_k} \nonumber \\
A^{(2)}_{k,j}&=&
\sum\limits_{i=1}^{N}
 \biggl(\fr{\omega+ f_0+\beta y}{H}\biggr)_i  \spm{\jac(p_i,p_j),p_k}
\eeqr

The scalar products with the operator $B$  in \rf{mateqpj} is  developed in a similar way. From \rf{B} we get
\beqr
\spm{B(\psi,\omega)\fr{\delta H}{H},p_k} &=& 
\spm{\jac(\psi, \fr{\omega+ f_0+\beta y}{H}\fr{\delta H}{H} ),p_k} +\nonumber \\
  &+&\spm{ \jac(\fr{\omega+ f_0+\beta y}{H},\biggl(\nabla \fr{1}{H}\nabla\biggr)^{-1} \biggl(  \nabla \fr{1}{H}\fr{\delta H}{H}  \nabla \psi\biggr)),p_k}
 \label{btmp}
\eeqr
Using \rf{develfun} we get for the first Jacobian 
\beq
\spm{\jac(\psi, \fr{\omega+ f_0+\beta y}{H}\fr{\delta H}{H} ),p_k}  = 
\sum\limits_{i=1}^{N}\sum\limits_{j=1}^{N} \psi_{i}(t)\biggl(\fr{\omega+ f_0+\beta y}{H}\biggr)_j \biggl(\fr{\delta H}{H}\biggr)_j \spm{\jac(p_i,p_j),p_k} \label{b1}
\eeq
and for the second one 
\beq
\spm{\jac(\fr{\omega+ f_0+\beta y}{H},\eta),p_k} =
\sum\limits_{i=1}^{N}\sum\limits_{j=1}^{N} \biggl(\fr{\omega+ f_0+\beta y}{H}\biggr)_i \eta_j \spm{\jac(p_i,p_j),p_k}
\label{b2} 
\eeq
where $\eta$ is determined from the equation
\beqr 
\biggl(\nabla \fr{1}{H}\nabla\biggr)^{-1} \biggl(  \nabla \fr{1}{H}\fr{\delta H}{H}  \nabla \psi\biggr) &=& \eta \nonumber \\
\biggl(\nabla \fr{1}{H}\nabla\biggr) \eta &=&\biggl(  \nabla \fr{1}{H}\fr{\delta H}{H}  \nabla \psi\biggr) 
\nonumber
\eeqr 
multiplying this equation by $p_k(x,y), \; \forall k=1 \ldots N$ and integrating by parts we get
\beqnn
\spm{\fr{1}{H}\nabla\eta ,\nabla p_k} = 
   \spm{  \fr{1}{H}\fr{\delta H}{H}\nabla \psi ,\nabla p_k}
\eeqnn 
The function $\eta$ is developed as liner combination of finite elements $\eta(x,y,t) = \sum\limits_{i=1}^{N} \eta_{i}(t) p_i(x,y)$ and we get  
 \beqnn
 \sum\limits_{i=1}^{N}\sum\limits_{j=1}^{N}\biggl(\fr{1}{H}\biggr)_i \eta_j \spm{p_i\nabla p_j,\nabla p_k} =\sum\limits_{i=1}^{N}\sum\limits_{j=1}^{N}\biggl(\fr{1}{H}\biggr)_i\biggl(\fr{\delta H}{H}\biggr)_i  \psi_j \spm{p_i\nabla p_j,\nabla p_k}
 \eeqnn
  or in matricial form 
 \beq
 {\cal H} \eta = {\cal P}  \fr{\delta H}{H} \mbox{ where } {\cal P}_{k,i} = 
\sum\limits_{j=1}^{N}\biggl(\fr{1}{H}\biggr)_i  \psi_j \spm{p_i\nabla p_j,\nabla p_k}
\label{matp}
\eeq
Combining  \rf{btmp}, \rf{b1}, \rf{b2} and \rf{matp} we get 
\beqr 
\spm{B(\psi,\omega)\fr{\delta H}{H} ,p_k} &=&  (B^{(1)}+B^{(2)}{\cal H}^{-1}{\cal P})\fr{\delta H}{H}  
\label{bmat} \\ 
B^{(1)}_{k,j}&=& \sum\limits_{i=1}^{N}  \psi_{i}(t)\biggl(\fr{\omega+ f_0+\beta y}{H}\biggr)_j  \spm{\jac(p_i,p_j),p_k}
\nonumber \\
B^{(2)}_{k,j}&=&\sum\limits_{i=1}^{N} \biggl(\fr{\omega+ f_0+\beta y}{H}\biggr)_i  \spm{\jac(p_i,p_j),p_k} \nonumber
\eeqr

Resuming all of the above, we get the finite element approximation of the equation \rf{mateq}
\beqr
{\cal M}  \der{\delta\omega(t)}{t} &=&(A^{(1)}+A^{(2)}{\cal H}^{-1}{\cal M}-\nu{\cal C} -\sigma{\cal M})\delta\omega + (B^{(1)}+B^{(2)}{\cal H}^{-1}{\cal P})\fr{\delta H}{H}  
\label{mateqfem} \\ 
A^{(1)}_{k,j}&=&\sum\limits_{i=1}^{N} \psi_{i}(t)\biggl(\fr{1}{H}\biggr)_j \spm{\jac(p_i,p_j),p_k} \nonumber \\
A^{(2)}_{k,j}&=&
\sum\limits_{i=1}^{N}
 \biggl(\fr{\omega+ f_0+\beta y}{H}\biggr)_i  \spm{\jac(p_i,p_j),p_k} 
 \nonumber\\
 B^{(1)}_{k,j}&=& \sum\limits_{i=1}^{N}  \psi_{i}(t)\biggl(\fr{\omega+ f_0+\beta y}{H}\biggr)_j  \spm{\jac(p_i,p_j),p_k}
\nonumber \\
B^{(2)}_{k,j}&=&\sum\limits_{i=1}^{N} \biggl(\fr{\omega+ f_0+\beta y}{H}\biggr)_i  \spm{\jac(p_i,p_j),p_k} \nonumber
\eeqr

\bibliography{/global/users/kazan/text/mybibl}
 
\mkpicstoend 

\end{document}